\documentclass[amsfonts, amssymb, amsmath, prl,twocolumn,showkeys, floatfix, twoside]{revtex4-2}
\usepackage[english]{babel}
\usepackage[utf8]{inputenc}
\usepackage[colorinlistoftodos, color=green!40, prependcaption]{todonotes}
\usepackage{amsthm}
\usepackage{mathtools}
\usepackage{physics}
\usepackage{xcolor}
\usepackage{graphicx}
\usepackage[left=23mm,right=13mm,top=35mm,columnsep=15pt]{geometry} 
\usepackage{adjustbox}
\usepackage{placeins}
\usepackage[T1]{fontenc}
\usepackage{lipsum}
\usepackage{csquotes}
\usepackage{soul}
\usepackage{xcolor}
\usepackage{bm}
\usepackage[pdftex, pdftitle={Article}, pdfauthor={Author}]{hyperref} % For hyperlinks in the PDF
\begin{document}
\title{Topological rigidity in twisted, elastic ribbons}

\author{Carlos E. Moguel-Lehmer}
    \email{cmoguell@syr.edu}
\author{Christian D. Santangelo}
    \email{cdsantan@syr.edu}
    \affiliation{Department of Physics, Syracuse University, Syracuse, New York 13244, USA}

\begin{abstract}
    Topology is an important determinant of the behavior of a great number of condensed-matter systems, but until recently has played a minor role in elasticity. We develop a theory for the deformations of a class of twisted non-Euclidean sheets which have a symmetry based on the celebrated Bonnet isometry. We show that non-orientability is an obstruction to realizing the symmetry globally, and induces a geometric phase that captures a memory analogous to a previously identified one in 2D metamaterials. However, we show that orientable ribbons can also obstruct realizing the symmetry globally. This new obstruction is mediated by how the unit normal vector winds around the centerline of the ribbon, and provides conditions for constructing soft modes of deformation compatible with the topology of multiply-twisted connected ribbons.
\end{abstract}

\keywords{Thin Sheets, Elasticity, Geometric Frustration.}

\maketitle

%%Introduction
%%
%%
%%
Recent work on topological phases in quantum matter \cite{hasan2010colloquium, prodan2009topological} has inspired new insights on classical rigidity, including the identification of topologically-protected edge modes in spring networks \cite{kane2014topological, lubensky2015phonons}, and the propagation of topological solitons \cite{chen2014nonlinear, lo2021topology}.
Though there are now powerful tools to produce thin but deformable structures of quite complex shapes and topologies, either directly \cite{xing2024shell} or by prescribing a metric \cite{klein2007shaping, momeni2017review}, how a shell's topology affects its rigidity remains underexplored.
The local nature of elasticity suggests that shell topology might play only an ancillary role, but topology and geometry are entwined and there is a well-known link between the rigidity of thin shells and their geometry \cite{ciarlet2005introduction, calladine1983theory, pogorelov1988bendings} pointing to a deeper connection.

It is well known, for example, that a spherical shell \cite{pogorelov1973} (or a generic simply-connected shell \cite{gluck2006almost}) is rigid unless a discontinuous ridge is formed \cite{gomez2016shallow}. In contrast, an elastic torus also has no boundaries but is anomalously soft along the curves where the Gaussian curvature changes sign \cite{audoly2002elastic, sun2022gol}. And ribbons twisted into M\"obius bands seem to inherit a number of interesting mechanical properties from their non-orientability, including gaps in the vibration spectrum for some polarizations \cite{nishiguchi2018phonon}, topological phononic edge modes \cite{chen2023topological}, continuous eversion \cite{nie2021light}, and a mechanical memory associated to non-orientable order \cite{bartolo2019topological, guo2023non}.

\begin{figure}[b]
    \centering
    \includegraphics[width=0.5\textwidth]{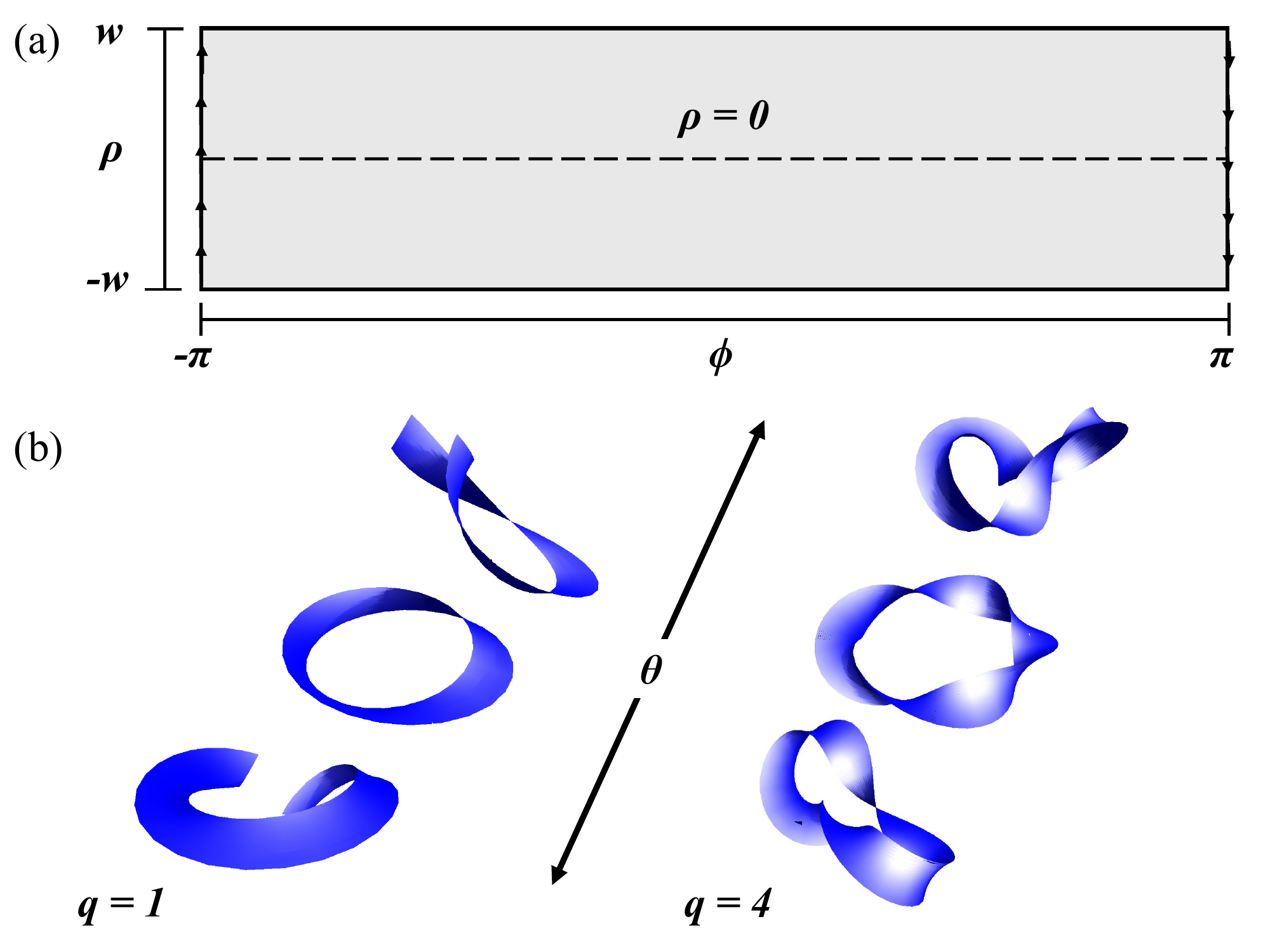}
    \caption{(a) Isothermal coordinate system for twisted ribbons $(\rho, \phi)$. For non-orientable ribbons, opposite sides are reflected as indicated by the arrows. (b) Two bent helicoids with $q=1$ and $q=4$ half twists. The Bonnet isometry, parameterized by the angle $\theta$, only preserves the topology for even integers $q > 2$.}
    \label{fig:surfaces}
\end{figure}

This letter explores the mechanics of twisted ribbons in the shape of the bent helicoids \cite{meeks2007bending}, a family of minimal surfaces characterized by an integer $q$ measuring the number of half-twists in the ribbon (Fig. \ref{fig:surfaces}). Odd $q$ ribbons are non-orientable; even $q$ ribbons are orientable. Because they are minimal surfaces, the ribbons have a continuous deformation, the Bonnet isometry, that preserves zero mean curvature without stretching. In the framework of incompatible elasticity \cite{efrati2009elastic}, in which the ribbons have the metric of the bent helicoids but a bending energy favoring flat geometries, the ribbons should exhibit an ultrasoft deformation \cite{efrati2009elastic, levin2016anomalously, arieli2024mechanical}. While true for even $q \ge 4$, we show that the connectivity of the $q=0$ and $q=2$ bent helicoids induces rigidity. However, ultrasoftness can be recovered by a suitable change in their geometry. Yet, all odd $q$ ribbons suffer a topological obstruction, because they are non-orientable, that preserves this topology-induced rigidity in all geometries.

%We find that, when \textcolor{blue}{the ends of the ribbons are glued together, so the ribbons must stay connected}, this ultrasoft deformation only occurs for ribbons with even $q \ge 4$ (Fig. \ref{fig:surfaces}). %We interpret this in terms of a periodic or antiperiodic scalar function associated the rotation of the deformation.

%Remarkably, the shape instabilities of nonorientable soap films turn out to share some of the hallmarks of elastic ribbons despite fundamental differences in their underlying elastic energies \cite{goldstein2010soap, pesci2015instability, machon2016instabilities}.

%Twisted ribbons under certain types of stress can also exhibit folds, wrinkles and other geometric features in response to twisting \cite{starostin2007shape, chopin2013helicoids, pham2016cylindrical}.

We use a system of dimensionless isothermal coordinates $(\rho, \phi)$, where $\rho$ spans the interval $[ -w, w ]$ and $w$ is the ratio of width to the characteristic radius of curvature (Fig. \ref{fig:surfaces}a) and $\partial_1$ and $\partial_2$ is shorthand for the partial derivatives with respect to $\rho$ and $\phi$ respectively. For each $q$, the bent helicoid $\mathbf{X}_q(\rho,\phi)$ is a minimal surface -- it has zero mean curvature -- and has the symmetry $\mathbf{X}_q(\rho,\phi + 2\pi) = \mathbf{X}_q\left( (-1)^q\rho,\phi \right)$, implying they are connected along $\phi$. Their metrics are isothermal,
\begin{equation}
    \label{metric bent helicoids}
    \bar{g}_{ij} = \partial_i\mathbf{X}_q \cdot \partial_j \mathbf{X}_q = \Omega_q(\rho, \phi) \delta_{ij},
\end{equation}
with $\Omega_q(\rho, \phi + 2\pi) = \Omega_q\left((-1)^q\rho, \phi\right)$.
Explicit formulas for the embeddings of the bent helicoids are cumbersome \cite{meeks2007bending, lopez2018explicit, machon2016instabilities} but are reproduced in the Supplemental Material (SM) \cite{Note1}.

In incompatible elasticity, elastic deformations
%are expressed by the shape of the membrane midsurface, $\mathbf{X}(\rho,\phi)$ \cite{efrati2009elastic}, by comparing its 
are measured by comparing the induced metric, $g_{i j} = \partial_i \mathbf{X} \cdot \partial_j \mathbf{X}$, to a prescribed metric $\bar{g}_{i j}$, and its induced curvature $b_{i j} = \hat{\mathbf{N}} \cdot \partial_i \partial_j \mathbf{X}$ where $\hat{\mathbf{N}}$ is the unit normal, to a prescribed curvature $\bar{b}_{i j}$. The prescribed metric and curvature represent the lateral distances and local curvatures that would make the sheet stress-free. %The induced metric and curvature represent similar quantities, but after a deformation occurs.
When $\bar{g} \neq g$, the sheet experiences in-plane stretching; when $\bar{b} \neq b$, it bends. We choose the prescribed metric to be compatible with the bent helicoid geometry of Eq.~(\ref{metric bent helicoids}) but $\bar{b}_{i j} = 0$.
In the absence of forces, the elastic energy has two terms, $E = E_s + E_b$. The stretching energy, $E_s$, can be written in terms of a strain tensor $\gamma_{i j} = (g_{i j} - \bar{g}_{i j})/2$ and the prescribed metric $\bar{g}_{i j}$ \cite{efrati2009elastic}. Under isometric deformations, the bending energy reduces to
\begin{equation}\label{eq:bending}
    E_b = \frac{\kappa B}{2} \int dA ~ H^2,
\end{equation}
where $H = g^{i j} b_{i j}/2$ is the mean curvature, $dA = d\rho d\phi \sqrt{\textrm{det}~\bar{g}_{i j}}$, $B$ is the bulk modulus of the sheet, and $\kappa$ is a dimensionless bending rigidity. Notice that $E_b$ is invariant under $H \rightarrow -H$ so it is defined for both orientable and non-orientable ribbons. The form of Eq. (\ref{eq:bending})
%, and particularly the absence of Gaussian curvature,%
reflects the bending energy of our numerical simulations. While the bending energy requires contracting the second fundamental form with the actual metric, it is unclear how to simulate it in a way independent of the discretization used, and some authors would argue that it introduces a spurious strain-bend coupling \cite{vitral2023dilation}. Using the prescribed metric instead overcomes these issues and only differs by terms proportional to the strain. Since we will ultimately be interested in isometries, however, the potential contributions we are neglecting should play only a minor role in determining shapes.

For a minimal surface, the Bonnet isometry
%is a symmetry of $E_s + E_b$, and 
can be defined in terms of an angle $\theta$ such that
\begin{equation}\label{eq:isom}
    \mathbf{X}_q(\rho, \phi; \theta) = \cos \theta ~ \mathbf{X}_q(\rho, \phi) + \sin \theta ~\mathbf{X}_q^C(\rho,\phi),
\end{equation}
where $\mathbf{X}_q^C(\rho,\phi)$, called the conjugate minimal surface to $\mathbf{X}_q$, is defined by the condition $\mathbf{X}_q(\rho, \phi;\pi/2) = \mathbf{X}_q^C(\rho,\phi)$ \cite{colding2011course, dierkes2010minimal}. The second fundamental form is $b_{i j} \cos \theta + b_{i j}^C \sin \theta$, where $b_{i j}^C$ is the second fundamental form of the conjugate minimal surface. The principal curvatures of $\mathbf{X}_q^C$ are rotated by $\pi/4$, so the Bonnet isometry induces a global rotation of the principle curvature directions by $\theta/2$ (see SM \footnote{See Supplemental Material at [ ], for analytical calculations and numerical details.}).
Though this symmetry can be weakly broken by boundary layers, minimal non-Euclidean plates still show anomalously soft elasticity under realistic conditions \cite{levin2016anomalously, arieli2024mechanical}. In disconnected ribbons, the Bonnet isometry also generates interesting topological phenomena such as fractional, solitonic excitations \cite{sun2021fractional}.

When a minimal surface is ``simply-connected'' -- so every closed loop can be continuously shrunk to a point \cite{munkres2000topology} -- the Bonnet isometry is the unique family of isometries (up to translations and rotations) preserving zero mean curvature \cite{nitsche1989lectures, schwarz1972gesammelte}. Our ribbons are not simply-connected, but a unique Bonnet isometry must exist on any simply-connected patch. Hence it must exist and be unique were we to cut the ribbons (see also SM \cite{Note1}). The question remains, however, when is there an obstruction to a global Bonnet isometry without cutting?

To identify which ribbons are disconnected by the Bonnet isometry, we consider their Bj\"orling problem: the minimal surface is constructed from a prescribed base curve, $\boldsymbol{\chi}(s)$, and unit normal vector, $\hat{\mathbf{N}}_q(s)$, perpendicular to it \cite{lopez2018explicit}. As we show in the SM \cite{Note1}, the displacement between the ends of the centerline of any ribbon $\mathbf{X}_q$ satisfies $\Delta\mathbf{X}_q(\theta) = \sin \theta ~ \mathbf{c}_q$ where
\begin{equation}
\label{eq:Centerline ribbon}
    \mathbf{c}_q = \int_0^{2\pi} \mathrm{ds} ~ \hat{\mathbf{N}}_q(s) \times \boldsymbol{\chi}'(s),
\end{equation}
where $s$ is arc length along the base curve. The vanishing of this integral is a necessary and sufficient condition for a minimal ribbon to remain connected. One can check that $q = 0$ and $q=2$ ribbons do not due to a resonance in the Fourier modes of $\hat{\mathbf{N}}_q$ and $\boldsymbol{\chi}'$ but $\mathbf{c}_q = 0$ for every other even $q > 2$. When $q$ is odd, the symmetry of the conjugate surfaces along $\rho=0$ satisfies \cite{Note1}
\begin{equation}
 \label{Conjugate surface Symmetry}
     \mathbf{X}_q^C(0, \phi + 2\pi)  = - \mathbf{X}_q^C(0, \phi) + \mathbf{c}_q.
\end{equation}
If $\mathbf{X}^C_q$ was closed, $\mathbf{X}_q^C(0, \phi + 2\pi) = \mathbf{X}_q^C(0, \phi)$ but this would imply $\mathbf{X}_q^C(0,\phi) = \mathbf{c}_q/2$, which is a contradiction. Hence, we expect only the even $q > 2$ ribbons to exhibit ultrasoftness.

%The Bonnet isometry always exists on simply-connected minimal surfaces, and in disconnected ribbons it can generate interesting topological phenomena such as fractional, solitonic excitations \cite{sun2021fractional}. Indeed, as is apparent by plotting the deformations (Fig. \ref{fig:surfaces}b), when the ribbons are closed, the isometry does not always preserve the ribbon's connectivity. 

%Instead, the bent helicoids satisfy the symmetry
%\begin{equation}
%\label{eq:symmetryX}
%    \mathbf{X}_q(\rho, \phi + 2 \pi; \theta) = \mathbf{X}_q[ (-1)^{q} \rho, \phi ; (-1)^{q} \theta ] + \Delta \mathbf{X}_q,
%\end{equation}
%where $\Delta \mathbf{X}_0 = -2 \pi \hat{z}$ and $\Delta \mathbf{X}_2 = \pi \hat{x}$ but is zero otherwise (see SM \cite{Note1}).
%It is natural then to distinguish three classes of bent helicoids: those with odd $q$ (non-orientable), those with even $q > 2$ (orientable), and the two special cases $q=0$ and $2$. We expect only the orientable class to exhibit ultrasoftness.
\begin{figure}[t]
   \centering
   \includegraphics[width=0.485\textwidth]{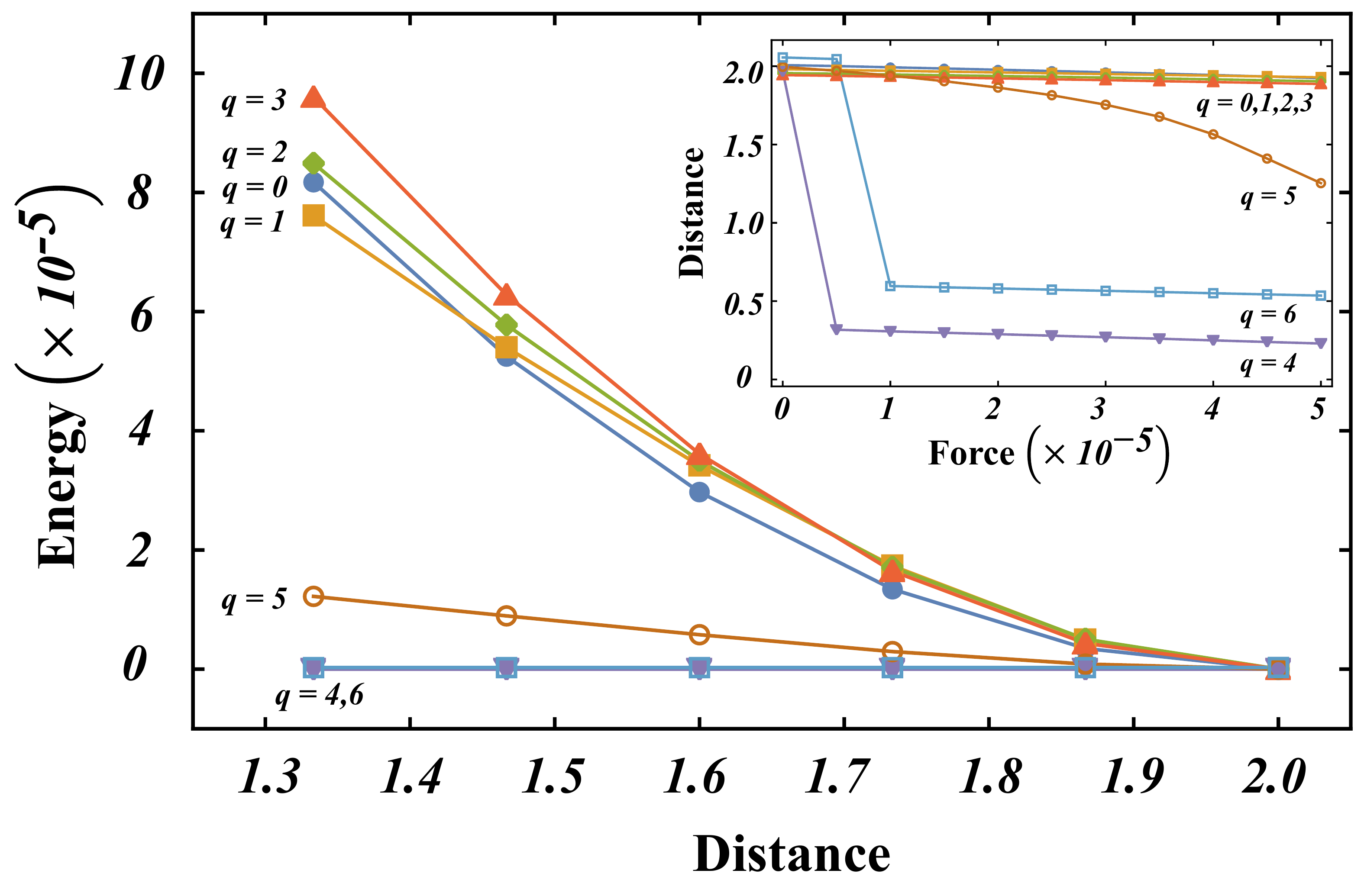}
   \caption{Energy as a function of the distance between antipodal points $(0,\pm \pi/2)$. In the continuum, the rest distance between antipodal points is at $2.0$ in the plot. There is almost no energetic cost to deform $q = 4$ and $q = 6$ compared to the other ribbons ($5520$ vertices). \emph{Inset}: Distance between antipodal points as a function of the dimensionless applied force, $f_0$ ($10,250$ vertices).}
   \label{fig:distance}
\end{figure}

To test this, we performed numerical minimization of a discretized bent helicoid on a triangular lattice with two different resolutions (see SM \cite{Note1}). We used a Seung-Nelson stretching energy, $E_{st} = \sum_{ij} a_{ij} (l_{i j}/\bar{l}_{i j} - 1)^2$ where the sum is over edges $(i,j)$ joining vertex $i$ to vertex $j$, $l_{ij}$ is the edge length, $\bar{l}_{i j}$ is the prescribed edge length before deformation, and $a_{ij}$ is the area of the sheet associated to each edge. We neglect the possibility of self-intersection. For this model, the Poisson ratio is estimated as $\nu = 1/3$ \cite{seung1988defects}. For the bending energy, we first locally correct the orientation of the faces around each vertex so that each vertex is surrounded by oriented faces, then compute the vertex discrete mean curvature \cite{sullivan2006curvature}. Finally, we apply a force pinching together the vertices closest to $(\rho,\phi) = (0, \pm \pi/2)$, associated with vertices $I$ and $J$ on the discrete lattice, by adding a term $f_0 B R l_{IJ}$ to the energy, where $R$ is the radius of the centerline and $f_0$ is dimensionless.

Unfortunately, the Bonnet isometry of the continuum is not a symmetry of any discrete surface (see SM \cite{Note1}). Nevertheless, Fig. \ref{fig:distance} clearly shows that the $q=4$ and $q=6$ ribbons are extremely soft to loading compared to the rest. Moreover, under cyclic loading, $q=4,6$ exhibit hysteresis while the response of the other ribbons is elastic (SM \cite{Note1}). Although $q = 6$ seems to achieve the smallest allowed distance by the continuum Bonnet isometry, we believe $q = 4$ does not due to discretization effects. Nevertheless, both deformations are in good agreement with the expected result from the Bonnet isometry (see \cite{Note1}). Note that we expect the distinction between different numbers of half-twists to disappear when $q$ is large, suggesting larger $q$ engenders a softer response. This is consistent with the relative softness for $q = 5$. All this is in sharp contrast to the extremely soft deformations expected of all ribbons that have been cut \cite{levin2016anomalously}. 

\begin{figure}[t]
   \centering
   \includegraphics[width=0.485\textwidth]{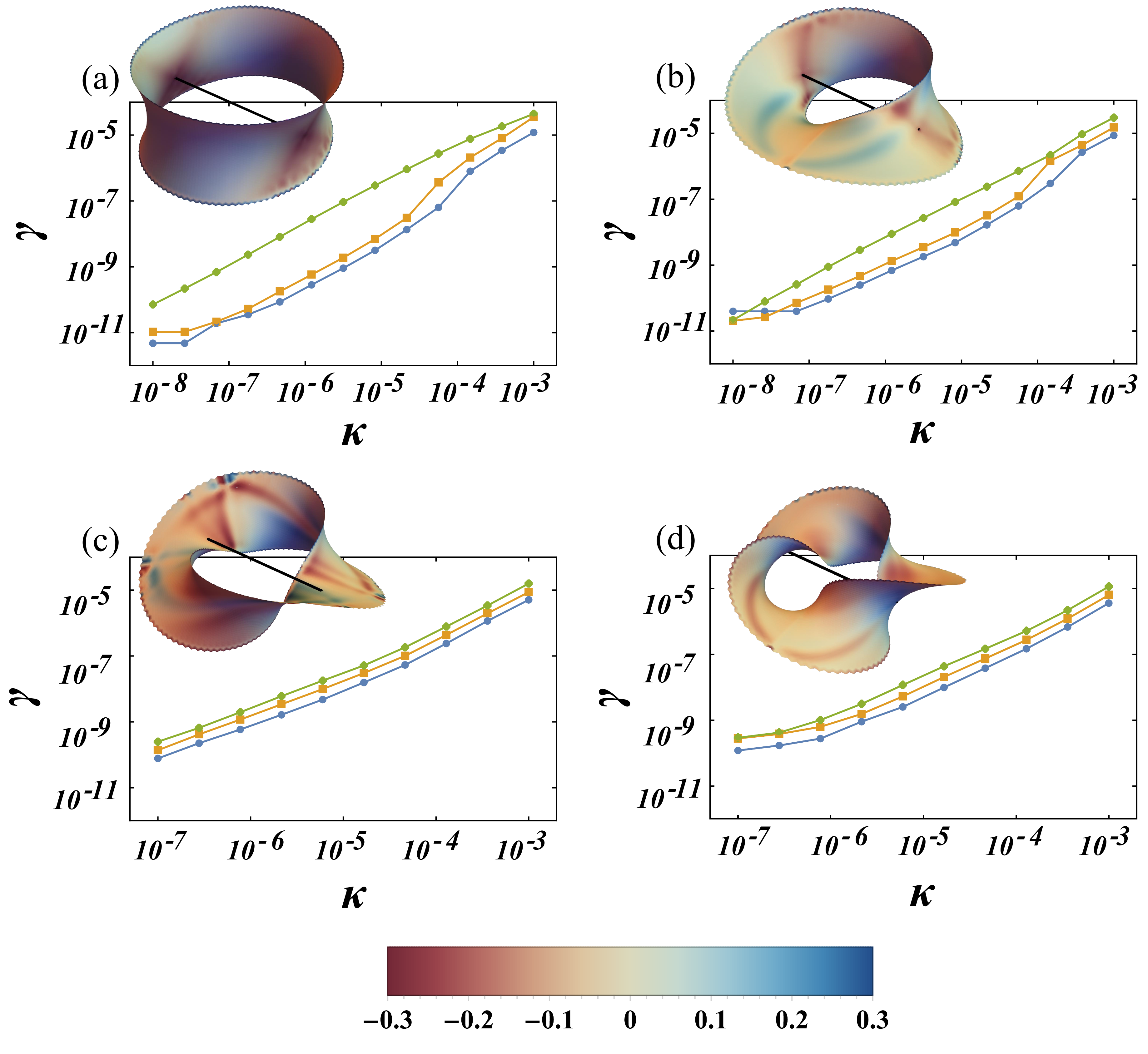}
   \caption{Mean strain as a function of bending rigidity $\kappa$ at fixed force ratio $f_0/\kappa$ for (a) $q = 0$, (b) $q = 1$, (c) $q = 2$, and (d) $q = 3$ bent helicoids. \emph{Diamond markers}: $f_0/\kappa = 1.0$, \emph{square markers}: $f_0/\kappa = 0.7$, \emph{circle markers}: $f_0/\kappa = 0.5$. The shading of the surfaces and colorbar represent the distribution of mean curvature. Ribbons exhibit small strains as the bending rigidity decreases.}
   \label{fig:sim}
\end{figure}

Fig. \ref{fig:sim} shows the mean strain, $\gamma = l_{i j}/\bar{l}_{i j} - 1$, as a function of the bending rigidity when the force $f_0/\kappa$ is held fixed, showing that $\gamma$ decreases dramatically as $\kappa$ decreases, consistent with an observed plateau in the final distance between antipodal points as $\kappa \rightarrow 0$ (SM \cite{Note1}). This justifies our use of isometries and Eq.~(\ref{eq:bending}).

%Fig. \ref{fig:sim}a-b shows the distance between the antipodal points as a function of the bending rigidity for $q=0$ and $q=1$ when the force $f_0/\kappa$ is held fixed. As $\kappa \rightarrow 0$, this distance reaches a plateau, suggesting that the stretching energy is subdominant and that the distance is selected by balancing bending stresses with the applied force. This is further corroborated by the mean strain $l_{i j}/\bar{l}_{i j} - 1$, \textcolor{blue}{ which decreases dramatically as $\kappa$ decreases (Fig. \ref{fig:sim}c-d). This justifies our use of isometries and Eq.~(\ref{eq:bending}).}

\begin{figure}[h!]
    \centering
    \includegraphics[width=1\linewidth]{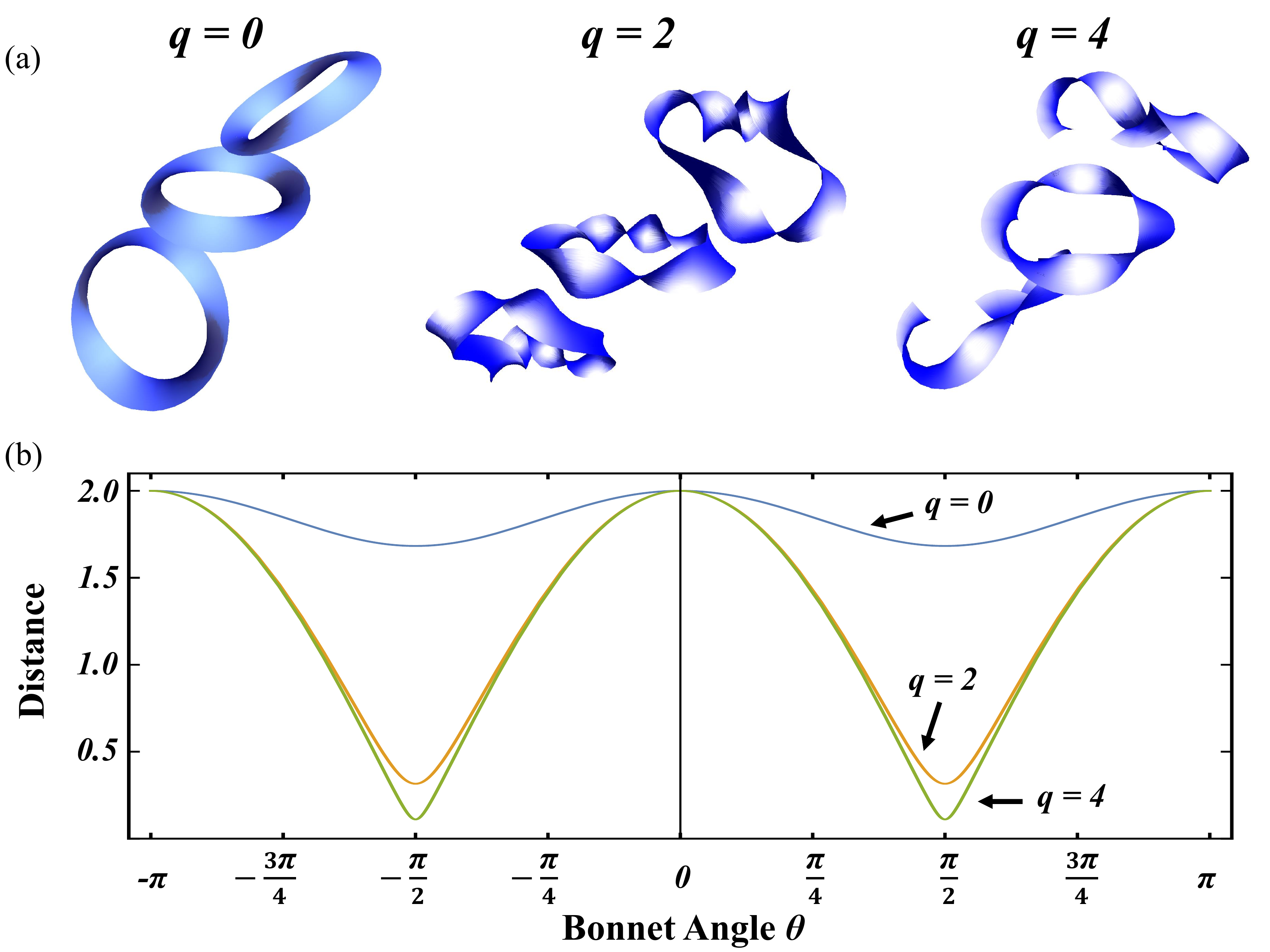}
    \caption{(a) Orientable minimal ribbons with the same half-twists as the $q = 0$ (left), $q = 2$ (middle), and $q = 4$ (right) bent helicoids. In contrast with the bent helicoids, $q = 0, 2$ ribbons are expected to be ultrasoft but $q = 4$ ribbon elastic if we glued the ends. (b) Distance between antipodal points $(0,\pm \pi/2)$ as a function of the Bonnet angle $\theta$, showing the ribbons are not undergoing a rigid body motion.}
    \label{fig:4}
\end{figure}

%All non-orientable minimal surfaces are elastic.
The topological rigidity of the bent helicoids raises a question: can the ribbon geometry be changed (while maintaining $H=0$) to restore the Bonnet isometry as a symmetry? We therefore search for a minimal ribbon with $q$ half-twists specified as a Bj\"orling problem. We define a circular base curve $\boldsymbol{\chi}(\phi) = \mathrm{R} (\sin \phi, \cos \phi, 0)^T$ of radius $\mathrm{R}$ and a unit normal vector $\hat{\mathbf{N}}(\phi)$ perpendicular to it that winds $q/2$ times as $\phi$ goes from $0$ to $2 \pi$. Then
\begin{equation}
    \mathbf{X}(\rho,\phi) = \textrm{Re} \left[ \boldsymbol{\chi}(z) + i \int dz ~ \hat{\mathbf{N}}(z) \times \boldsymbol{\chi}'(z) \right]
\end{equation}
is a minimal surface, where $z = \phi + i \rho$ \cite{lopez2018explicit}.
A Bonnet isometry preventing the ribbon from tearing would then require $\int_0^{2 \pi R} \mathrm{ds} ~ \hat{\mathbf{N}}(s) \times \boldsymbol{\chi}'(s) = 0$ where $s = R \phi$ is the arc length along the center line.
We achieve this with the choice
\begin{eqnarray}
    \hat{\mathbf{N}}(\phi) &=& -\sin \phi \sin \left(q \phi/2 + a \sin \phi\right) \hat{\mathbf{x}} \nonumber \\
   & &  -\cos \phi \sin \left(q \phi/2 + a \sin \phi\right) \hat{\mathbf{y}} \\
   & & + \cos \left( q \phi/2 + a \sin \phi\right) \hat{\mathbf{z}}, \nonumber
\end{eqnarray}
where $a$ is any zero of the Bessel function $J_{q/2}$ for both $q=0$ and $q=2$. Similarly, it is possible to perturb the bent helicoids for $q>2$ to frustrate the global Bonnet isometry (Fig. \ref{fig:4}).

%\textcolor{blue}{%with the half-twists of the orientable bent helicoids by prescribing a different family of unit normal vectors, $\hat{\mathbf{N}}_n$. 
%The displacement of the centerline of the ribbon under the Bonnet isometry is still given by Eq.~(\ref{eq:Centerline ribbon}). The vanishing of the integral $\int_0^{2\pi} \hat{\mathbf{N}}(s) \times \boldsymbol{\chi}'(s) \ \mathrm{ds}$ defines a torus knot whose center of mass must be at the center of the torus. The torus knot is parameterized as $\boldsymbol{w}(s) = \boldsymbol{\chi}(s) + \hat{\mathbf{N}}(s) \times \boldsymbol{\chi}'(s)$, where $s$ is a parameter along the base curve of the bent helicoids, $\boldsymbol{\chi}(s) = R(\sin s, \cos s, 0)^T$ where $R$ is a constant. Note that $\int_0^{2\pi} \hat{\mathbf{N}}(s) \times \boldsymbol{\chi}'(s) \ \mathrm{ds} = 0$ if and only if $\int_0^{2\pi} \boldsymbol{w}(s)~\mathrm{ds} =0$. The orientable bent helicoids, with $\hat{\mathbf{N}} = \hat{\mathbf{N}}_q$, define $\left(1, \frac{q}{2}\right)$ torus knots satisfying the contraint when $q > 2$. However, these knots can be deformed, while preserving their windings around the torus, so that $\hat{\mathbf{N}} = \hat{\mathbf{N}}_n$. For these new minimal surfaces, the $(1, 0)$ and $(1, 1)$ torus knots maintain the surface connected during the deformation, while the other $(1, n)$ knots frustrate the Bonnet isometry (Fig.~\ref{fig:4}). Hence, the rigidity of orientable minimal ribbons is not entirely topological. Explicit parameterizations of deformed torus knots are found in the Supplementary Material (SM) \cite{Note1}.}

On any non-orientable minimal surface, however, this turns out to be impossible. Non-orientability guarantees that there is a non-contractible loop $\alpha$ on the surface $\mathbf{X}$ over which, after parallel transporting around the loop, the unit normal to the surface, $\hat{\mathbf{N}}$ satisfies $\hat{\mathbf{N}} \mapsto -\hat{\mathbf{N}}$ \cite{do2016differential}. By solving a Bj\"orling problem with $\alpha$ as the prescribed base curve and $\hat{\mathbf{N}}$ as the prescribed unit normal, we can construct a subset of $\mathbf{X}$ parameterized in isothermal coordinates $(\rho, \phi)$ such that $\partial_\phi \mathbf{X} = \alpha'(\phi)$ and $\partial_\rho \mathbf{X}$ is a tangent vector orthogonal to the centerline \cite{nitsche1989lectures}. These tangent vectors satisfy the symmetries $\partial_\rho\mathbf{X}(0, \phi + 2\pi) = -\partial_\rho\mathbf{X}(0, \phi)$ and $\partial_\phi \mathbf{X}(0, \phi + 2\pi) = \partial_\phi \mathbf{X}(0, \phi)$ along the base curve by construction. The restrictions of Eqs.~(\ref{eq:Centerline ribbon}) and (\ref{Conjugate surface Symmetry}) to the base curve imply the newly constructed ribbon, and hence the whole surface, disconnects under the Bonnet isometry. We show in the SM that this is independent of the choice of non-contractible loop $\alpha$ \cite{Note1}.

%What are the mechanical consequences of non-orientable rigidity in ribbons?
Having identified non-orientable minimal ribbons as the only generically frustrated ones, we now want to understand how they deform under load. To that end, we consider an arbitrary isometry, $\delta \mathbf{X} = u^i \partial_i \mathbf{X}_q + \zeta \hat{\mathbf{N}}_q$, where
\begin{equation}\label{eq:isometrystrain}
    D_i u_j + D_j u_i - 2 \zeta b_{i j} = 0,
\end{equation}
and the covariant derivative $D_i$, the metric $g_{i j}$, and the curvature $b_{i j}$ are those of $\mathbf{X}_q$. We also define $b^C_{i j}$ to be the curvature of the conjugate surface $\mathbf{X}_q^C$. Since $g^{i j} b_{i j} = 0$ and $b_{ij} b_C^{i j} = 0$, Eq. (\ref{eq:isometrystrain}) reduces to the system of equations $D_i u^i = 0$,  $b^{i j} D_i u_j + 2 K \zeta = 0$, and  $b_C^{i j} D_i u_j = 0$,
where $K$ is the Gaussian curvature of $\mathbf{X}_q$. The first equation represents the vanishing of areal strains and the other two are associated with shear strains within the sheet.

A generic in-plane deformation on an orientable surface with a boundary is given by the Hodge-Morrey-Friedrich decomposition (HMF), $u_i = \partial_i \eta + \epsilon_{i}^{~j} \partial_j \psi + h_i$ where $\epsilon_{i j}$ is the Levi-Civita symbol, $D_i h_i = 0$ and $\epsilon^{i j} D_i h_i = 0$, $\eta$ is constant on the boundary, and the tangent component of $h_i$ vanishes on the boundary \cite{SchwarzHodge1995, bhatia2012helmholtz}. Therefore, $\eta$ is a constant, which we set to zero, and $h_i = 0$. Hence, we obtain
\begin{eqnarray}
\label{Isometry eqs}
    b^{i j} D_i D_j \psi &=& 0,  \\
    b_C^{i j} D_i D_j \psi &=& -2 K \zeta. \nonumber
\end{eqnarray}
We can ascribe physical meaning to $\psi$ by considering the rotation pseudoscalar of the isometry, $\chi = \epsilon^{i j} D_i u_j = - \nabla^2_{\bar{g}} \psi$, where $\nabla^2_{\bar{g}}$ is the Laplace-Beltrami operator on $\mathbf{X}_q$.

%Non-orientable order.
On non-orientable ribbons, we use the HMF decomposition on the orientable double cover ($-2 \pi < \phi \le 2 \pi$) but apply the additional condition that $\psi$ be compatible with the ribbon symmetry.
From symmetry of the ribbons when $\phi \rightarrow \pi + 2 \pi$, 
%Since the deformation $\delta \mathbf{X}[ (-1)^q \rho,\phi+2\pi] = \delta \mathbf{X}(\rho,\phi)$ in order for the ribbon to remain closed,
we obtain
\begin{equation}
    \psi \left( (-1)^q \rho,\phi+2 \pi \right) = (-1)^q \psi \left( \rho, \phi \right).
\end{equation}
The rotation $\chi$ satisfies the same symmetry. As we show in the SM \cite{Note1}, the Bonnet isometry has $\chi = -2$ because the principal curvatures rotate by a constant angle everywhere on the surface; this is inconsistent with the required antisymmetry of odd $q$. Indeed, the only constant $\chi$ is $\chi = 0$, and there must be a point of zero rotation somewhere on the surface \cite{nakahara2018geometry}.

The antisymmetric rotation $\chi$ thus emerges as a natural signature of non-orientable order in 2D sheets, analogous to non-orientable quasi-1D ribbons, for which Refs. \cite{bartolo2019topological, guo2023non} introduced an auxiliary degree of freedom associated with the location of a topologically-protected point of zero deformation. Here, the nodal point in 1D becomes a nodal curve in 2D along which $\chi(\rho,\phi) = 0$, and the antiperiodicity of $\chi$ provides a memory of the deformation history of the ribbon through a $\mathbb{Z}_2$ geometric phase \cite{berry1990budden}. For example, let $\delta \mathbf{X}(\rho,\phi + \phi_0) \ne 0$ describe a family of infinitesimal deformations with parameter $\phi_0$. Taking $\phi_0$ continuously from $0$ to $2 \pi$ takes $\delta \mathbf{X}(\rho,\phi)$ to $\delta \mathbf{X}(\rho,\phi+2\pi) = -\delta \mathbf{X}(-\rho, \phi)$. Therefore, any shape for which $-\delta \mathbf{X}(-\rho,\phi) \ne \delta \mathbf{X}(\rho,\phi)$ does not return to itself.
%{We expand $\psi(\rho,\phi) = \sum_n \psi_n(\phi) \rho^n$. Since $\partial_\rho \mathbf{X}(\rho,\phi) \cdot \delta \mathbf{X}(\rho,\phi) = \partial_\phi \psi(\rho,\phi) = \psi_0'(\phi) + \rho \psi_1'(\phi) + \mathcal{O}(\rho^2)$, any deformation for which $\psi_1'(\phi) \ne 0$ will not return to itself as $\phi$ goes from $0$ to $2 \pi$. } 

%\textcolor{blue}{Although intractable for general $q$, we can analytically solve Eqs.~(\ref{Isometry eqs}) in the case of $q = 0$ and obtain good agreement with numerical results. To keep the focus on the topology, we relegate this explicit calculation to the SM \cite{Note1}.}

%Summary.
In summary, we have studied the deformations of orientable and non-orientable non-Euclidean ribbons based on the bent helicoids. Despite the existence of a soft mode of the elastic energy, we identify two mechanisms that prevent the symmetry from being realized globally, one geometric and one topological having to do with non-orientability. This demonstrates that global constraints can induce a topological rigidity in non-simply connected shells that cannot exist when they are simply connected. We also show that non-orientable order is realized by an antiperiodic scalar function associated with local rotation \cite{guo2023non} and encodes a $\mathbb{Z}_2$ geometric phase. These results pave the way toward understanding how topology determines elastic response in continuum elastic sheets.

%\textcolor{blue}{We expect ultrasoft connected ribbons, which we construct by solving a Bj\"orling problem, to show mechanical properties independent of their prescribed base curve and unit normal since they all deform along the same soft mode, but we leave their in-depth study for future work.}

\begin{acknowledgements}
    We are indebted to Denis Bartolo, Corentin Coulais, Sabetta Matsumoto, and Gareth Alexander for insightful comments. We are particularly thankful to the anonymous referees for inspiring new results. We acknowledge funding through NSF CMMI-2247095.
\end{acknowledgements}

\bibliography{references}

\begin{thebibliography}{7}%
\makeatletter
\providecommand \@ifxundefined [1]{%
 \@ifx{#1\undefined}
}%
\providecommand \@ifnum [1]{%
 \ifnum #1\expandafter \@firstoftwo
 \else \expandafter \@secondoftwo
 \fi
}%
\providecommand \@ifx [1]{%
 \ifx #1\expandafter \@firstoftwo
 \else \expandafter \@secondoftwo
 \fi
}%
\providecommand \natexlab [1]{#1}%
\providecommand \enquote  [1]{``#1''}%
\providecommand \bibnamefont  [1]{#1}%
\providecommand \bibfnamefont [1]{#1}%
\providecommand \citenamefont [1]{#1}%
\providecommand \href@noop [0]{\@secondoftwo}%
\providecommand \href [0]{\begingroup \@sanitize@url \@href}%
\providecommand \@href[1]{\@@startlink{#1}\@@href}%
\providecommand \@@href[1]{\endgroup#1\@@endlink}%
\providecommand \@sanitize@url [0]{\catcode `\\12\catcode `\$12\catcode `\&12\catcode `\#12\catcode `\^12\catcode `\_12\catcode `\%12\relax}%
\providecommand \@@startlink[1]{}%
\providecommand \@@endlink[0]{}%
\providecommand \url  [0]{\begingroup\@sanitize@url \@url }%
\providecommand \@url [1]{\endgroup\@href {#1}{\urlprefix }}%
\providecommand \urlprefix  [0]{URL }%
\providecommand \Eprint [0]{\href }%
\providecommand \doibase [0]{http://dx.doi.org/}%
\providecommand \selectlanguage [0]{\@gobble}%
\providecommand \bibinfo  [0]{\@secondoftwo}%
\providecommand \bibfield  [0]{\@secondoftwo}%
\providecommand \translation [1]{[#1]}%
\providecommand \BibitemOpen [0]{}%
\providecommand \bibitemStop [0]{}%
\providecommand \bibitemNoStop [0]{.\EOS\space}%
\providecommand \EOS [0]{\spacefactor3000\relax}%
\providecommand \BibitemShut  [1]{\csname bibitem#1\endcsname}%
\let\auto@bib@innerbib\@empty
%</preamble>
\bibitem [{\citenamefont {Dierkes}\ \emph {et~al.}(2010)\citenamefont {Dierkes}, \citenamefont {Hildebrandt}, \citenamefont {Sauvigny}, \citenamefont {Dierkes}, \citenamefont {Hildebrandt},\ and\ \citenamefont {Sauvigny}}]{dierkes2010minimal}%
  \BibitemOpen
  \bibfield  {author} {\bibinfo {author} {\bibfnamefont {U.}~\bibnamefont {Dierkes}}, \bibinfo {author} {\bibfnamefont {S.}~\bibnamefont {Hildebrandt}}, \bibinfo {author} {\bibfnamefont {F.}~\bibnamefont {Sauvigny}}, \bibinfo {author} {\bibfnamefont {U.}~\bibnamefont {Dierkes}}, \bibinfo {author} {\bibfnamefont {S.}~\bibnamefont {Hildebrandt}}, \ and\ \bibinfo {author} {\bibfnamefont {F.}~\bibnamefont {Sauvigny}},\ }\href@noop {} {\emph {\bibinfo {title} {Minimal surfaces}}}\ (\bibinfo  {publisher} {Springer},\ \bibinfo {year} {2010})\BibitemShut {NoStop}%
\bibitem [{\citenamefont {Nitsche}(1989)}]{nitsche1989lectures}%
  \BibitemOpen
  \bibfield  {author} {\bibinfo {author} {\bibfnamefont {J.}~\bibnamefont {Nitsche}},\ }\href@noop {} {\enquote {\bibinfo {title} {Lectures on minimal surfaces. vol. 1},}\ } (\bibinfo {year} {1989})\BibitemShut {NoStop}%
\bibitem [{\citenamefont {Schwarz}(1972)}]{schwarz1972gesammelte}%
  \BibitemOpen
  \bibfield  {author} {\bibinfo {author} {\bibfnamefont {H.~A.}\ \bibnamefont {Schwarz}},\ }\href@noop {} {\emph {\bibinfo {title} {Gesammelte mathematische abhandlungen}}},\ Vol.\ \bibinfo {volume} {260}\ (\bibinfo  {publisher} {American Mathematical Soc.},\ \bibinfo {year} {1972})\BibitemShut {NoStop}%
\bibitem [{\citenamefont {Kose}\ \emph {et~al.}(2011)\citenamefont {Kose}, \citenamefont {Toda},\ and\ \citenamefont {Aulisa}}]{kose2011solving}%
  \BibitemOpen
  \bibfield  {author} {\bibinfo {author} {\bibfnamefont {Z.}~\bibnamefont {Kose}}, \bibinfo {author} {\bibfnamefont {M.}~\bibnamefont {Toda}}, \ and\ \bibinfo {author} {\bibfnamefont {E.}~\bibnamefont {Aulisa}},\ }\href@noop {} {\bibfield  {journal} {\bibinfo  {journal} {Balkan Journal of Geometry and Its Applications}\ }\textbf {\bibinfo {volume} {16}},\ \bibinfo {pages} {70} (\bibinfo {year} {2011})}\BibitemShut {NoStop}%
\bibitem [{\citenamefont {Schwarz}(2006)}]{schwarz2006hodge}%
  \BibitemOpen
  \bibfield  {author} {\bibinfo {author} {\bibfnamefont {G.}~\bibnamefont {Schwarz}},\ }\href@noop {} {\emph {\bibinfo {title} {Hodge Decomposition-A method for solving boundary value problems}}}\ (\bibinfo  {publisher} {Springer},\ \bibinfo {year} {2006})\BibitemShut {NoStop}%
\bibitem [{\citenamefont {Meeks~III}\ and\ \citenamefont {Weber}(2007)}]{meeks2007bending}%
  \BibitemOpen
  \bibfield  {author} {\bibinfo {author} {\bibfnamefont {W.~H.}\ \bibnamefont {Meeks~III}}\ and\ \bibinfo {author} {\bibfnamefont {M.}~\bibnamefont {Weber}},\ }\href@noop {} {\bibfield  {journal} {\bibinfo  {journal} {Mathematische Annalen}\ }\textbf {\bibinfo {volume} {339}},\ \bibinfo {pages} {783} (\bibinfo {year} {2007})}\BibitemShut {NoStop}%
\bibitem [{\citenamefont {L{\'o}pez}\ and\ \citenamefont {Weber}(2018)}]{lopez2018explicit}%
  \BibitemOpen
  \bibfield  {author} {\bibinfo {author} {\bibfnamefont {R.}~\bibnamefont {L{\'o}pez}}\ and\ \bibinfo {author} {\bibfnamefont {M.}~\bibnamefont {Weber}},\ }\href@noop {} {\bibfield  {journal} {\bibinfo  {journal} {Michigan Mathematical Journal}\ }\textbf {\bibinfo {volume} {67}},\ \bibinfo {pages} {561} (\bibinfo {year} {2018})}\BibitemShut {NoStop}%
\end{thebibliography}%


\begin{thebibliography}{43}%
\makeatletter
\providecommand \@ifxundefined [1]{%
 \@ifx{#1\undefined}
}%
\providecommand \@ifnum [1]{%
 \ifnum #1\expandafter \@firstoftwo
 \else \expandafter \@secondoftwo
 \fi
}%
\providecommand \@ifx [1]{%
 \ifx #1\expandafter \@firstoftwo
 \else \expandafter \@secondoftwo
 \fi
}%
\providecommand \natexlab [1]{#1}%
\providecommand \enquote  [1]{``#1''}%
\providecommand \bibnamefont  [1]{#1}%
\providecommand \bibfnamefont [1]{#1}%
\providecommand \citenamefont [1]{#1}%
\providecommand \href@noop [0]{\@secondoftwo}%
\providecommand \href [0]{\begingroup \@sanitize@url \@href}%
\providecommand \@href[1]{\@@startlink{#1}\@@href}%
\providecommand \@@href[1]{\endgroup#1\@@endlink}%
\providecommand \@sanitize@url [0]{\catcode `\\12\catcode `\$12\catcode `\&12\catcode `\#12\catcode `\^12\catcode `\_12\catcode `\%12\relax}%
\providecommand \@@startlink[1]{}%
\providecommand \@@endlink[0]{}%
\providecommand \url  [0]{\begingroup\@sanitize@url \@url }%
\providecommand \@url [1]{\endgroup\@href {#1}{\urlprefix }}%
\providecommand \urlprefix  [0]{URL }%
\providecommand \Eprint [0]{\href }%
\providecommand \doibase [0]{http://dx.doi.org/}%
\providecommand \selectlanguage [0]{\@gobble}%
\providecommand \bibinfo  [0]{\@secondoftwo}%
\providecommand \bibfield  [0]{\@secondoftwo}%
\providecommand \translation [1]{[#1]}%
\providecommand \BibitemOpen [0]{}%
\providecommand \bibitemStop [0]{}%
\providecommand \bibitemNoStop [0]{.\EOS\space}%
\providecommand \EOS [0]{\spacefactor3000\relax}%
\providecommand \BibitemShut  [1]{\csname bibitem#1\endcsname}%
\let\auto@bib@innerbib\@empty
%</preamble>
\bibitem [{\citenamefont {Hasan}\ and\ \citenamefont {Kane}(2010)}]{hasan2010colloquium}%
  \BibitemOpen
  \bibfield  {author} {\bibinfo {author} {\bibfnamefont {M.~Z.}\ \bibnamefont {Hasan}}\ and\ \bibinfo {author} {\bibfnamefont {C.~L.}\ \bibnamefont {Kane}},\ }\href@noop {} {\bibfield  {journal} {\bibinfo  {journal} {Reviews of modern physics}\ }\textbf {\bibinfo {volume} {82}},\ \bibinfo {pages} {3045} (\bibinfo {year} {2010})}\BibitemShut {NoStop}%
\bibitem [{\citenamefont {Prodan}\ and\ \citenamefont {Prodan}(2009)}]{prodan2009topological}%
  \BibitemOpen
  \bibfield  {author} {\bibinfo {author} {\bibfnamefont {E.}~\bibnamefont {Prodan}}\ and\ \bibinfo {author} {\bibfnamefont {C.}~\bibnamefont {Prodan}},\ }\href@noop {} {\bibfield  {journal} {\bibinfo  {journal} {Physical review letters}\ }\textbf {\bibinfo {volume} {103}},\ \bibinfo {pages} {248101} (\bibinfo {year} {2009})}\BibitemShut {NoStop}%
\bibitem [{\citenamefont {Kane}\ and\ \citenamefont {Lubensky}(2014)}]{kane2014topological}%
  \BibitemOpen
  \bibfield  {author} {\bibinfo {author} {\bibfnamefont {C.~L.}\ \bibnamefont {Kane}}\ and\ \bibinfo {author} {\bibfnamefont {T.~C.}\ \bibnamefont {Lubensky}},\ }\href@noop {} {\bibfield  {journal} {\bibinfo  {journal} {Nature Physics}\ }\textbf {\bibinfo {volume} {10}},\ \bibinfo {pages} {39} (\bibinfo {year} {2014})}\BibitemShut {NoStop}%
\bibitem [{\citenamefont {Lubensky}\ \emph {et~al.}(2015)\citenamefont {Lubensky}, \citenamefont {Kane}, \citenamefont {Mao}, \citenamefont {Souslov},\ and\ \citenamefont {Sun}}]{lubensky2015phonons}%
  \BibitemOpen
  \bibfield  {author} {\bibinfo {author} {\bibfnamefont {T.}~\bibnamefont {Lubensky}}, \bibinfo {author} {\bibfnamefont {C.}~\bibnamefont {Kane}}, \bibinfo {author} {\bibfnamefont {X.}~\bibnamefont {Mao}}, \bibinfo {author} {\bibfnamefont {A.}~\bibnamefont {Souslov}}, \ and\ \bibinfo {author} {\bibfnamefont {K.}~\bibnamefont {Sun}},\ }\href@noop {} {\bibfield  {journal} {\bibinfo  {journal} {Reports on Progress in Physics}\ }\textbf {\bibinfo {volume} {78}},\ \bibinfo {pages} {073901} (\bibinfo {year} {2015})}\BibitemShut {NoStop}%
\bibitem [{\citenamefont {Chen}\ \emph {et~al.}(2014)\citenamefont {Chen}, \citenamefont {Upadhyaya},\ and\ \citenamefont {Vitelli}}]{chen2014nonlinear}%
  \BibitemOpen
  \bibfield  {author} {\bibinfo {author} {\bibfnamefont {B.~G.-g.}\ \bibnamefont {Chen}}, \bibinfo {author} {\bibfnamefont {N.}~\bibnamefont {Upadhyaya}}, \ and\ \bibinfo {author} {\bibfnamefont {V.}~\bibnamefont {Vitelli}},\ }\href@noop {} {\bibfield  {journal} {\bibinfo  {journal} {Proceedings of the National Academy of Sciences}\ }\textbf {\bibinfo {volume} {111}},\ \bibinfo {pages} {13004} (\bibinfo {year} {2014})}\BibitemShut {NoStop}%
\bibitem [{\citenamefont {Lo}\ \emph {et~al.}(2021)\citenamefont {Lo}, \citenamefont {Santangelo}, \citenamefont {Chen}, \citenamefont {Jian}, \citenamefont {Roychowdhury},\ and\ \citenamefont {Lawler}}]{lo2021topology}%
  \BibitemOpen
  \bibfield  {author} {\bibinfo {author} {\bibfnamefont {P.-W.}\ \bibnamefont {Lo}}, \bibinfo {author} {\bibfnamefont {C.~D.}\ \bibnamefont {Santangelo}}, \bibinfo {author} {\bibfnamefont {B.~G.-g.}\ \bibnamefont {Chen}}, \bibinfo {author} {\bibfnamefont {C.-M.}\ \bibnamefont {Jian}}, \bibinfo {author} {\bibfnamefont {K.}~\bibnamefont {Roychowdhury}}, \ and\ \bibinfo {author} {\bibfnamefont {M.~J.}\ \bibnamefont {Lawler}},\ }\href@noop {} {\bibfield  {journal} {\bibinfo  {journal} {Physical Review Letters}\ }\textbf {\bibinfo {volume} {127}},\ \bibinfo {pages} {076802} (\bibinfo {year} {2021})}\BibitemShut {NoStop}%
\bibitem [{\citenamefont {Xing}\ \emph {et~al.}(2024)\citenamefont {Xing}, \citenamefont {Wang}, \citenamefont {Lu}, \citenamefont {Sharf}, \citenamefont {Cohen-Or},\ and\ \citenamefont {Tu}}]{xing2024shell}%
  \BibitemOpen
  \bibfield  {author} {\bibinfo {author} {\bibfnamefont {Y.}~\bibnamefont {Xing}}, \bibinfo {author} {\bibfnamefont {X.}~\bibnamefont {Wang}}, \bibinfo {author} {\bibfnamefont {L.}~\bibnamefont {Lu}}, \bibinfo {author} {\bibfnamefont {A.}~\bibnamefont {Sharf}}, \bibinfo {author} {\bibfnamefont {D.}~\bibnamefont {Cohen-Or}}, \ and\ \bibinfo {author} {\bibfnamefont {C.}~\bibnamefont {Tu}},\ }\href@noop {} {\bibfield  {journal} {\bibinfo  {journal} {Computational Visual Media}\ ,\ \bibinfo {pages} {1}} (\bibinfo {year} {2024})}\BibitemShut {NoStop}%
\bibitem [{\citenamefont {Klein}\ \emph {et~al.}(2007)\citenamefont {Klein}, \citenamefont {Efrati},\ and\ \citenamefont {Sharon}}]{klein2007shaping}%
  \BibitemOpen
  \bibfield  {author} {\bibinfo {author} {\bibfnamefont {Y.}~\bibnamefont {Klein}}, \bibinfo {author} {\bibfnamefont {E.}~\bibnamefont {Efrati}}, \ and\ \bibinfo {author} {\bibfnamefont {E.}~\bibnamefont {Sharon}},\ }\href@noop {} {\bibfield  {journal} {\bibinfo  {journal} {Science}\ }\textbf {\bibinfo {volume} {315}},\ \bibinfo {pages} {1116} (\bibinfo {year} {2007})}\BibitemShut {NoStop}%
\bibitem [{\citenamefont {Momeni}\ \emph {et~al.}(2017)\citenamefont {Momeni}, \citenamefont {Liu}, \citenamefont {Ni} \emph {et~al.}}]{momeni2017review}%
  \BibitemOpen
  \bibfield  {author} {\bibinfo {author} {\bibfnamefont {F.}~\bibnamefont {Momeni}}, \bibinfo {author} {\bibfnamefont {X.}~\bibnamefont {Liu}}, \bibinfo {author} {\bibfnamefont {J.}~\bibnamefont {Ni}},  \emph {et~al.},\ }\href@noop {} {\bibfield  {journal} {\bibinfo  {journal} {Materials \& design}\ }\textbf {\bibinfo {volume} {122}},\ \bibinfo {pages} {42} (\bibinfo {year} {2017})}\BibitemShut {NoStop}%
\bibitem [{\citenamefont {Ciarlet}(2005)}]{ciarlet2005introduction}%
  \BibitemOpen
  \bibfield  {author} {\bibinfo {author} {\bibfnamefont {P.~G.}\ \bibnamefont {Ciarlet}},\ }\href@noop {} {\bibfield  {journal} {\bibinfo  {journal} {Journal of elasticity}\ }\textbf {\bibinfo {volume} {78}},\ \bibinfo {pages} {1} (\bibinfo {year} {2005})}\BibitemShut {NoStop}%
\bibitem [{\citenamefont {Calladine}(1983)}]{calladine1983theory}%
  \BibitemOpen
  \bibfield  {author} {\bibinfo {author} {\bibfnamefont {C.~R.}\ \bibnamefont {Calladine}},\ }\href@noop {} {\emph {\bibinfo {title} {Theory of shell structures}}}\ (\bibinfo  {publisher} {Cambridge university press},\ \bibinfo {year} {1983})\BibitemShut {NoStop}%
\bibitem [{\citenamefont {Pogorelov}(1988)}]{pogorelov1988bendings}%
  \BibitemOpen
  \bibfield  {author} {\bibinfo {author} {\bibfnamefont {A.~V.}\ \bibnamefont {Pogorelov}},\ }\href@noop {} {\emph {\bibinfo {title} {Bendings of surfaces and stability of shells}}},\ Vol.~\bibinfo {volume} {72}\ (\bibinfo  {publisher} {American Mathematical Soc.},\ \bibinfo {year} {1988})\BibitemShut {NoStop}%
\bibitem [{\citenamefont {Pogorelov}(1973)}]{pogorelov1973}%
  \BibitemOpen
  \bibfield  {author} {\bibinfo {author} {\bibfnamefont {A.~V.}\ \bibnamefont {Pogorelov}},\ }\href@noop {} {\emph {\bibinfo {title} {Extrinsic geometry of convex surfaces}}},\ Vol.~\bibinfo {volume} {35}\ (\bibinfo  {publisher} {American Mathematical Soc.},\ \bibinfo {year} {1973})\BibitemShut {NoStop}%
\bibitem [{\citenamefont {Gluck}(2006)}]{gluck2006almost}%
  \BibitemOpen
  \bibfield  {author} {\bibinfo {author} {\bibfnamefont {H.}~\bibnamefont {Gluck}},\ }in\ \href@noop {} {\emph {\bibinfo {booktitle} {Geometric Topology: Proceedings of the Geometric Topology Conference held at Park City, Utah, February 19--22, 1974}}}\ (\bibinfo {organization} {Springer},\ \bibinfo {year} {2006})\ pp.\ \bibinfo {pages} {225--239}\BibitemShut {NoStop}%
\bibitem [{\citenamefont {Gomez}\ \emph {et~al.}(2016)\citenamefont {Gomez}, \citenamefont {Moulton},\ and\ \citenamefont {Vella}}]{gomez2016shallow}%
  \BibitemOpen
  \bibfield  {author} {\bibinfo {author} {\bibfnamefont {M.}~\bibnamefont {Gomez}}, \bibinfo {author} {\bibfnamefont {D.~E.}\ \bibnamefont {Moulton}}, \ and\ \bibinfo {author} {\bibfnamefont {D.}~\bibnamefont {Vella}},\ }\href@noop {} {\bibfield  {journal} {\bibinfo  {journal} {Proceedings of the Royal Society A: Mathematical, Physical and Engineering Sciences}\ }\textbf {\bibinfo {volume} {472}},\ \bibinfo {pages} {20150732} (\bibinfo {year} {2016})}\BibitemShut {NoStop}%
\bibitem [{\citenamefont {Audoly}\ and\ \citenamefont {Pomeau}(2002)}]{audoly2002elastic}%
  \BibitemOpen
  \bibfield  {author} {\bibinfo {author} {\bibfnamefont {B.}~\bibnamefont {Audoly}}\ and\ \bibinfo {author} {\bibfnamefont {Y.}~\bibnamefont {Pomeau}},\ }\href@noop {} {\bibfield  {journal} {\bibinfo  {journal} {Comptes Rendus M{\'e}canique}\ }\textbf {\bibinfo {volume} {330}},\ \bibinfo {pages} {425} (\bibinfo {year} {2002})}\BibitemShut {NoStop}%
\bibitem [{\citenamefont {Sun}(2022)}]{sun2022gol}%
  \BibitemOpen
  \bibfield  {author} {\bibinfo {author} {\bibfnamefont {B.}~\bibnamefont {Sun}},\ }\href@noop {} {\bibfield  {journal} {\bibinfo  {journal} {Thin-Walled Structures}\ }\textbf {\bibinfo {volume} {171}},\ \bibinfo {pages} {108718} (\bibinfo {year} {2022})}\BibitemShut {NoStop}%
\bibitem [{\citenamefont {Nishiguchi}\ and\ \citenamefont {Wybourne}(2018)}]{nishiguchi2018phonon}%
  \BibitemOpen
  \bibfield  {author} {\bibinfo {author} {\bibfnamefont {N.}~\bibnamefont {Nishiguchi}}\ and\ \bibinfo {author} {\bibfnamefont {M.~N.}\ \bibnamefont {Wybourne}},\ }\href@noop {} {\bibfield  {journal} {\bibinfo  {journal} {Journal of Physics Communications}\ }\textbf {\bibinfo {volume} {2}},\ \bibinfo {pages} {085002} (\bibinfo {year} {2018})}\BibitemShut {NoStop}%
\bibitem [{\citenamefont {Chen}\ \emph {et~al.}(2023)\citenamefont {Chen}, \citenamefont {Chen}, \citenamefont {Wu}, \citenamefont {Chen},\ and\ \citenamefont {Zheng}}]{chen2023topological}%
  \BibitemOpen
  \bibfield  {author} {\bibinfo {author} {\bibfnamefont {Y.}~\bibnamefont {Chen}}, \bibinfo {author} {\bibfnamefont {J.-H.}\ \bibnamefont {Chen}}, \bibinfo {author} {\bibfnamefont {F.}~\bibnamefont {Wu}}, \bibinfo {author} {\bibfnamefont {H.}~\bibnamefont {Chen}}, \ and\ \bibinfo {author} {\bibfnamefont {Z.-H.}\ \bibnamefont {Zheng}},\ }\href@noop {} {\bibfield  {journal} {\bibinfo  {journal} {Results in Physics}\ }\textbf {\bibinfo {volume} {46}},\ \bibinfo {pages} {106322} (\bibinfo {year} {2023})}\BibitemShut {NoStop}%
\bibitem [{\citenamefont {Nie}\ \emph {et~al.}(2021)\citenamefont {Nie}, \citenamefont {Zuo}, \citenamefont {Wang}, \citenamefont {Huang}, \citenamefont {Chen}, \citenamefont {Liu},\ and\ \citenamefont {Yang}}]{nie2021light}%
  \BibitemOpen
  \bibfield  {author} {\bibinfo {author} {\bibfnamefont {Z.-Z.}\ \bibnamefont {Nie}}, \bibinfo {author} {\bibfnamefont {B.}~\bibnamefont {Zuo}}, \bibinfo {author} {\bibfnamefont {M.}~\bibnamefont {Wang}}, \bibinfo {author} {\bibfnamefont {S.}~\bibnamefont {Huang}}, \bibinfo {author} {\bibfnamefont {X.-M.}\ \bibnamefont {Chen}}, \bibinfo {author} {\bibfnamefont {Z.-Y.}\ \bibnamefont {Liu}}, \ and\ \bibinfo {author} {\bibfnamefont {H.}~\bibnamefont {Yang}},\ }\href@noop {} {\bibfield  {journal} {\bibinfo  {journal} {Nature Communications}\ }\textbf {\bibinfo {volume} {12}},\ \bibinfo {pages} {2334} (\bibinfo {year} {2021})}\BibitemShut {NoStop}%
\bibitem [{\citenamefont {Bartolo}\ and\ \citenamefont {Carpentier}(2019)}]{bartolo2019topological}%
  \BibitemOpen
  \bibfield  {author} {\bibinfo {author} {\bibfnamefont {D.}~\bibnamefont {Bartolo}}\ and\ \bibinfo {author} {\bibfnamefont {D.}~\bibnamefont {Carpentier}},\ }\href@noop {} {\bibfield  {journal} {\bibinfo  {journal} {Physical Review X}\ }\textbf {\bibinfo {volume} {9}},\ \bibinfo {pages} {041058} (\bibinfo {year} {2019})}\BibitemShut {NoStop}%
\bibitem [{\citenamefont {Guo}\ \emph {et~al.}(2023)\citenamefont {Guo}, \citenamefont {Guzm{\'a}n}, \citenamefont {Carpentier}, \citenamefont {Bartolo},\ and\ \citenamefont {Coulais}}]{guo2023non}%
  \BibitemOpen
  \bibfield  {author} {\bibinfo {author} {\bibfnamefont {X.}~\bibnamefont {Guo}}, \bibinfo {author} {\bibfnamefont {M.}~\bibnamefont {Guzm{\'a}n}}, \bibinfo {author} {\bibfnamefont {D.}~\bibnamefont {Carpentier}}, \bibinfo {author} {\bibfnamefont {D.}~\bibnamefont {Bartolo}}, \ and\ \bibinfo {author} {\bibfnamefont {C.}~\bibnamefont {Coulais}},\ }\href@noop {} {\bibfield  {journal} {\bibinfo  {journal} {Nature}\ }\textbf {\bibinfo {volume} {618}},\ \bibinfo {pages} {506} (\bibinfo {year} {2023})}\BibitemShut {NoStop}%
\bibitem [{\citenamefont {Meeks~III}\ and\ \citenamefont {Weber}(2007)}]{meeks2007bending}%
  \BibitemOpen
  \bibfield  {author} {\bibinfo {author} {\bibfnamefont {W.~H.}\ \bibnamefont {Meeks~III}}\ and\ \bibinfo {author} {\bibfnamefont {M.}~\bibnamefont {Weber}},\ }\href@noop {} {\bibfield  {journal} {\bibinfo  {journal} {Mathematische Annalen}\ }\textbf {\bibinfo {volume} {339}},\ \bibinfo {pages} {783} (\bibinfo {year} {2007})}\BibitemShut {NoStop}%
\bibitem [{\citenamefont {Efrati}\ \emph {et~al.}(2009)\citenamefont {Efrati}, \citenamefont {Sharon},\ and\ \citenamefont {Kupferman}}]{efrati2009elastic}%
  \BibitemOpen
  \bibfield  {author} {\bibinfo {author} {\bibfnamefont {E.}~\bibnamefont {Efrati}}, \bibinfo {author} {\bibfnamefont {E.}~\bibnamefont {Sharon}}, \ and\ \bibinfo {author} {\bibfnamefont {R.}~\bibnamefont {Kupferman}},\ }\href@noop {} {\bibfield  {journal} {\bibinfo  {journal} {Journal of the Mechanics and Physics of Solids}\ }\textbf {\bibinfo {volume} {57}},\ \bibinfo {pages} {762} (\bibinfo {year} {2009})}\BibitemShut {NoStop}%
\bibitem [{\citenamefont {Levin}\ and\ \citenamefont {Sharon}(2016)}]{levin2016anomalously}%
  \BibitemOpen
  \bibfield  {author} {\bibinfo {author} {\bibfnamefont {I.}~\bibnamefont {Levin}}\ and\ \bibinfo {author} {\bibfnamefont {E.}~\bibnamefont {Sharon}},\ }\href@noop {} {\bibfield  {journal} {\bibinfo  {journal} {Physical review letters}\ }\textbf {\bibinfo {volume} {116}},\ \bibinfo {pages} {035502} (\bibinfo {year} {2016})}\BibitemShut {NoStop}%
\bibitem [{\citenamefont {Arieli}\ \emph {et~al.}(2024)\citenamefont {Arieli}, \citenamefont {Moshe},\ and\ \citenamefont {Sharon}}]{arieli2024mechanical}%
  \BibitemOpen
  \bibfield  {author} {\bibinfo {author} {\bibfnamefont {M.}~\bibnamefont {Arieli}}, \bibinfo {author} {\bibfnamefont {M.}~\bibnamefont {Moshe}}, \ and\ \bibinfo {author} {\bibfnamefont {E.}~\bibnamefont {Sharon}},\ }\href@noop {} {\bibfield  {journal} {\bibinfo  {journal} {Soft Matter}\ }\textbf {\bibinfo {volume} {20}},\ \bibinfo {pages} {4414} (\bibinfo {year} {2024})}\BibitemShut {NoStop}%
\bibitem [{\citenamefont {L{\'o}pez}\ and\ \citenamefont {Weber}(2018)}]{lopez2018explicit}%
  \BibitemOpen
  \bibfield  {author} {\bibinfo {author} {\bibfnamefont {R.}~\bibnamefont {L{\'o}pez}}\ and\ \bibinfo {author} {\bibfnamefont {M.}~\bibnamefont {Weber}},\ }\href@noop {} {\bibfield  {journal} {\bibinfo  {journal} {Michigan Mathematical Journal}\ }\textbf {\bibinfo {volume} {67}},\ \bibinfo {pages} {561} (\bibinfo {year} {2018})}\BibitemShut {NoStop}%
\bibitem [{\citenamefont {Machon}\ \emph {et~al.}(2016)\citenamefont {Machon}, \citenamefont {Alexander}, \citenamefont {Goldstein},\ and\ \citenamefont {Pesci}}]{machon2016instabilities}%
  \BibitemOpen
  \bibfield  {author} {\bibinfo {author} {\bibfnamefont {T.}~\bibnamefont {Machon}}, \bibinfo {author} {\bibfnamefont {G.~P.}\ \bibnamefont {Alexander}}, \bibinfo {author} {\bibfnamefont {R.~E.}\ \bibnamefont {Goldstein}}, \ and\ \bibinfo {author} {\bibfnamefont {A.~I.}\ \bibnamefont {Pesci}},\ }\href@noop {} {\bibfield  {journal} {\bibinfo  {journal} {Physical Review Letters}\ }\textbf {\bibinfo {volume} {117}},\ \bibinfo {pages} {017801} (\bibinfo {year} {2016})}\BibitemShut {NoStop}%
\bibitem [{Note1()}]{Note1}%
  \BibitemOpen
  \bibinfo {note} {See Supplemental Material at [ ], for analytical calculations and numerical details.}\BibitemShut {Stop}%
\bibitem [{\citenamefont {Vitral}\ and\ \citenamefont {Hanna}(2023)}]{vitral2023dilation}%
  \BibitemOpen
  \bibfield  {author} {\bibinfo {author} {\bibfnamefont {E.}~\bibnamefont {Vitral}}\ and\ \bibinfo {author} {\bibfnamefont {J.}~\bibnamefont {Hanna}},\ }\href@noop {} {\bibfield  {journal} {\bibinfo  {journal} {Journal of Elasticity}\ }\textbf {\bibinfo {volume} {153}},\ \bibinfo {pages} {571} (\bibinfo {year} {2023})}\BibitemShut {NoStop}%
\bibitem [{\citenamefont {Colding}\ and\ \citenamefont {Minicozzi}(2011)}]{colding2011course}%
  \BibitemOpen
  \bibfield  {author} {\bibinfo {author} {\bibfnamefont {T.~H.}\ \bibnamefont {Colding}}\ and\ \bibinfo {author} {\bibfnamefont {W.~P.}\ \bibnamefont {Minicozzi}},\ }\href@noop {} {\emph {\bibinfo {title} {A course in minimal surfaces}}},\ Vol.\ \bibinfo {volume} {121}\ (\bibinfo  {publisher} {American Mathematical Soc.},\ \bibinfo {year} {2011})\BibitemShut {NoStop}%
\bibitem [{\citenamefont {Dierkes}\ \emph {et~al.}(2010)\citenamefont {Dierkes}, \citenamefont {Hildebrandt}, \citenamefont {Sauvigny}, \citenamefont {Dierkes}, \citenamefont {Hildebrandt},\ and\ \citenamefont {Sauvigny}}]{dierkes2010minimal}%
  \BibitemOpen
  \bibfield  {author} {\bibinfo {author} {\bibfnamefont {U.}~\bibnamefont {Dierkes}}, \bibinfo {author} {\bibfnamefont {S.}~\bibnamefont {Hildebrandt}}, \bibinfo {author} {\bibfnamefont {F.}~\bibnamefont {Sauvigny}}, \bibinfo {author} {\bibfnamefont {U.}~\bibnamefont {Dierkes}}, \bibinfo {author} {\bibfnamefont {S.}~\bibnamefont {Hildebrandt}}, \ and\ \bibinfo {author} {\bibfnamefont {F.}~\bibnamefont {Sauvigny}},\ }\href@noop {} {\emph {\bibinfo {title} {Minimal surfaces}}}\ (\bibinfo  {publisher} {Springer},\ \bibinfo {year} {2010})\BibitemShut {NoStop}%
\bibitem [{\citenamefont {Sun}\ and\ \citenamefont {Mao}(2021)}]{sun2021fractional}%
  \BibitemOpen
  \bibfield  {author} {\bibinfo {author} {\bibfnamefont {K.}~\bibnamefont {Sun}}\ and\ \bibinfo {author} {\bibfnamefont {X.}~\bibnamefont {Mao}},\ }\href@noop {} {\bibfield  {journal} {\bibinfo  {journal} {Physical Review Letters}\ }\textbf {\bibinfo {volume} {127}},\ \bibinfo {pages} {098001} (\bibinfo {year} {2021})}\BibitemShut {NoStop}%
\bibitem [{\citenamefont {Munkres}(2000)}]{munkres2000topology}%
  \BibitemOpen
  \bibfield  {author} {\bibinfo {author} {\bibfnamefont {J.}~\bibnamefont {Munkres}},\ }\href {https://books.google.com/books?id=XjoZAQAAIAAJ} {\emph {\bibinfo {title} {Topology}}},\ Featured Titles for Topology\ (\bibinfo  {publisher} {Prentice Hall, Incorporated},\ \bibinfo {year} {2000})\BibitemShut {NoStop}%
\bibitem [{\citenamefont {Nitsche}(1989)}]{nitsche1989lectures}%
  \BibitemOpen
  \bibfield  {author} {\bibinfo {author} {\bibfnamefont {J.}~\bibnamefont {Nitsche}},\ }\href@noop {} {\enquote {\bibinfo {title} {Lectures on minimal surfaces. vol. 1},}\ } (\bibinfo {year} {1989})\BibitemShut {NoStop}%
\bibitem [{\citenamefont {Schwarz}(1972)}]{schwarz1972gesammelte}%
  \BibitemOpen
  \bibfield  {author} {\bibinfo {author} {\bibfnamefont {H.~A.}\ \bibnamefont {Schwarz}},\ }\href@noop {} {\emph {\bibinfo {title} {Gesammelte mathematische abhandlungen}}},\ Vol.\ \bibinfo {volume} {260}\ (\bibinfo  {publisher} {American Mathematical Soc.},\ \bibinfo {year} {1972})\BibitemShut {NoStop}%
\bibitem [{\citenamefont {Seung}\ and\ \citenamefont {Nelson}(1988)}]{seung1988defects}%
  \BibitemOpen
  \bibfield  {author} {\bibinfo {author} {\bibfnamefont {H.~S.}\ \bibnamefont {Seung}}\ and\ \bibinfo {author} {\bibfnamefont {D.~R.}\ \bibnamefont {Nelson}},\ }\href@noop {} {\bibfield  {journal} {\bibinfo  {journal} {Physical Review A}\ }\textbf {\bibinfo {volume} {38}},\ \bibinfo {pages} {1005} (\bibinfo {year} {1988})}\BibitemShut {NoStop}%
\bibitem [{\citenamefont {Sullivan}(2006)}]{sullivan2006curvature}%
  \BibitemOpen
  \bibfield  {author} {\bibinfo {author} {\bibfnamefont {J.~M.}\ \bibnamefont {Sullivan}},\ }in\ \href@noop {} {\emph {\bibinfo {booktitle} {ACM SIGGRAPH 2006 Courses}}}\ (\bibinfo {year} {2006})\ pp.\ \bibinfo {pages} {10--13}\BibitemShut {NoStop}%
\bibitem [{\citenamefont {Do~Carmo}(2016)}]{do2016differential}%
  \BibitemOpen
  \bibfield  {author} {\bibinfo {author} {\bibfnamefont {M.~P.}\ \bibnamefont {Do~Carmo}},\ }\href@noop {} {\emph {\bibinfo {title} {Differential geometry of curves and surfaces: revised and updated second edition}}}\ (\bibinfo  {publisher} {Courier Dover Publications},\ \bibinfo {year} {2016})\BibitemShut {NoStop}%
\bibitem [{\citenamefont {Schwarz}(1995)}]{SchwarzHodge1995}%
  \BibitemOpen
  \bibfield  {author} {\bibinfo {author} {\bibfnamefont {G.}~\bibnamefont {Schwarz}},\ }\href@noop {} {\emph {\bibinfo {title} {Hodge Decomposition – A Method for Solving Boundary Value Problems}}}\ (\bibinfo  {publisher} {Springer Berlin, Heidelberg},\ \bibinfo {year} {1995})\BibitemShut {NoStop}%
\bibitem [{\citenamefont {Bhatia}\ \emph {et~al.}(2012)\citenamefont {Bhatia}, \citenamefont {Norgard}, \citenamefont {Pascucci},\ and\ \citenamefont {Bremer}}]{bhatia2012helmholtz}%
  \BibitemOpen
  \bibfield  {author} {\bibinfo {author} {\bibfnamefont {H.}~\bibnamefont {Bhatia}}, \bibinfo {author} {\bibfnamefont {G.}~\bibnamefont {Norgard}}, \bibinfo {author} {\bibfnamefont {V.}~\bibnamefont {Pascucci}}, \ and\ \bibinfo {author} {\bibfnamefont {P.-T.}\ \bibnamefont {Bremer}},\ }\href@noop {} {\bibfield  {journal} {\bibinfo  {journal} {IEEE Transactions on visualization and computer graphics}\ }\textbf {\bibinfo {volume} {19}},\ \bibinfo {pages} {1386} (\bibinfo {year} {2012})}\BibitemShut {NoStop}%
\bibitem [{\citenamefont {Nakahara}(2018)}]{nakahara2018geometry}%
  \BibitemOpen
  \bibfield  {author} {\bibinfo {author} {\bibfnamefont {M.}~\bibnamefont {Nakahara}},\ }\href@noop {} {\emph {\bibinfo {title} {Geometry, topology and physics}}}\ (\bibinfo  {publisher} {CRC press},\ \bibinfo {year} {2018})\BibitemShut {NoStop}%
\bibitem [{\citenamefont {Berry}(1990)}]{berry1990budden}%
  \BibitemOpen
  \bibfield  {author} {\bibinfo {author} {\bibfnamefont {M.~V.}\ \bibnamefont {Berry}},\ }\href@noop {} {\bibfield  {journal} {\bibinfo  {journal} {Proceedings of the Royal Society of London. Series A: Mathematical and Physical Sciences}\ }\textbf {\bibinfo {volume} {431}},\ \bibinfo {pages} {531} (\bibinfo {year} {1990})}\BibitemShut {NoStop}%
\end{thebibliography}%

\end{document}

\title{Supplementary Material: Topological rigidity in twisted, elastic sheets}

\author{Carlos E. Moguel-Lehmer}
    \email{cmoguell@syr.edu}
\author{Christian D. Santangelo}
    \email{cdsantan@syr.edu}
    \affiliation{Department of Physics, Syracuse University, Syracuse, New York 13244, USA}

\maketitle

\tableofcontents

\pagebreak 

\section{Minimal surfaces and their isometries}

\subsection{The associate family of a minimal surface}
All minimal surfaces are part of a one-parameter family of isometric minimal surfaces called the asssociate family. The associate family of a minimal surface can be written in the form $\mathbf{X} = \cos\theta ~ \mathbf{X}_0 + \sin \theta ~ \mathbf{X}_C$ where $\theta$ is the Bonnet angle. In isothermal coordinates, where $ds^2 = \Omega(x^1, x^2) [ (dx^1)^2 + (dx^2)^2 ]$, the surfaces $\mathbf{X}_0$ and $\mathbf{X}_C$ satisfy the Cauchy-Riemann equations \cite{dierkes2010minimal},
\begin{equation}
\label{Cauchy-Riemann}
\partial_1\mathbf{X}_0 = \partial_2\mathbf{X}_C, \quad \partial_2\mathbf{X}_0 = -\partial_1\mathbf{X}_C
\end{equation}
Hence, $\nabla^2 X_0 = \nabla^2 X_C = 0$, where $\nabla^2 = \partial_1^2 + \partial_2^2$, the normal vectors of $\mathbf{X}_0$ and $\mathbf{X_C}$ are equal for each $(\rho,\phi)$, and the normal vector for any constant $\theta$ is,
\begin{equation}
\label{Normal}
\hat{\mathbf{N}} = \frac{1}{\sqrt{\det {g}}}\left(\partial_1\mathbf{X} \times \partial_2
\mathbf{X}\right) = \cos^2\theta ~ \hat{\mathbf{N}}_0 + \sin^2\theta ~ \hat{\mathbf{N}}_0 = \hat{\mathbf{N}}_0
\end{equation}
where $\hat{\mathbf{N}}_0$ is the normal unit vector to $\mathbf{X}_0$. A direct computation using eq.~(\ref{Normal}) and the Cauchy-Riemann equations shows that the second fundamental form is given by,
\begin{equation}
\label{Curvature tensor Bonnet}
b_{i j} = \cos\theta ~b^0_{i j} + \sin\theta ~ b^C_{i j}
\end{equation}
where $b^0$ and $b^C$ are the second fundamental forms of $\mathbf{X}_0$ and $\mathbf{X}_C$, respectively.

The Bonnet isometry rotates the principal curvatures of the second fundamental form by $\theta/2$ \cite{nitsche1989lectures}. For completeness, we will show this below. The eigenvectors, $e_\pm^i$, of the second fundamental form (\ref{Curvature tensor Bonnet}) satisfy the covariant eigenvalue equation in isothermal coordinates,
\begin{equation}
    \left(b_{i j} \pm c ~ \Omega\delta_{ij}\right) e_\pm^j = 0,
\end{equation}
and the corresponding eigenvalues are denoted $\pm c$.
%The normalized eigenvector corresponding to the eigenvalue $c$ is one of the principal directions in the actual configuration and can be written
Thus, in the principal frame of $\mathbf{X}_0$, 
\begin{equation}
\label{eigenvector curvature tensor}
e_+(\theta) = \left(e_+^1, e_+^2\right) = \sqrt{\frac{\Omega}{2\left(1 + \cos \theta\right)}} \left(1 + \cos \theta, \sin \theta\right).
\end{equation}
The principal direction, $e_+(0)$, in the reference configuration is
\begin{equation}
    e_+(0) = \sqrt{\Omega}\left(1,0\right).
\end{equation}
Thus, the angle, $\psi$, between $e_+(0)$ and $e_+(\theta)$ for general $\theta$ is given by,
\begin{equation}
\label{Angle of rotation}
\cos \psi = \frac{e_+^i(\theta) ~ {g}_{ij} e_+^j(0)}{|e_+(\theta)||e_+(0)|} = \sqrt{\frac{1 + \cos \theta}{2}} = \cos\left(\frac{\theta}{2}\right),
\end{equation}
where $|v| = \sqrt{ g_{i j} v^i v^j }$.
%We conclude the Bonnet isometry rotates the principal direction in the reference configuration by an angle $\theta/2$.

\subsection{Uniqueness of the Bonnet isometry}
Schwarz's theorem \cite{schwarz1972gesammelte, nitsche1989lectures} says that if a simply-connected minimal surface, $\mathrm{S}$, is isometric to another minimal surface, $\mathrm{S'}$, then $\mathrm{S'}$ a member of the associate family of $\mathrm{S}$. This identifies the Bonnet isometry as the unique isometry that preserves the mean curvature on minimal surfaces. Here, we provide an alternate way to arrive at the same conclusion that we think is enlightening.

Bonnet's theorem says that the first and second fundamental forms of a surface determine its shape up to Euclidean motions \cite{kose2011solving}. In an isothermal coordinate system, the Gauss-Codazzi-Mainardi system must satisfy the equations
\begin{eqnarray}
    b_{1 1} b_{2 2} - b_{1 2}^2 &=& - \Omega \frac{\nabla^2 \ln \Omega}{2}, \\
    \partial_2 b_{1 1} - \partial_1 b_{12} &=& b_{1 1} \frac{\partial_2 \Omega}{2 \Omega} + b_{22} \frac{\partial_2 \Omega}{2 \Omega}, \\
    \partial_2 b_{12} - \partial_1 b_{22} &=& -b_{11} \frac{\partial_1 \Omega}{2 \Omega} - b_{22} \frac{\partial_1 \Omega}{2 \Omega}.
\end{eqnarray}

When $H = 0$ as well, $b_{11} = -b_{22}$, and
\begin{eqnarray}
    b_{11}^2 + b_{12}^2 &=& \frac{1}{2} \Omega \nabla^2 \ln \Omega \\
    \partial_2 b_{11} - \partial_1 b_{12} &=& 0,  \\
    \partial_2 b_{12} - \partial_1 b_{22} &=& 0 .
\end{eqnarray}
Defining $f^2 \equiv (1/2) \Omega \nabla^2 \ln \Omega$ allows us to write $b_{11} = f \cos \alpha$, and $b_{12} = f \sin \alpha$. Then our equations reduce to a pair of equations,
\begin{eqnarray}\label{eq:psi}
    \begin{array}{rcl}
    \partial_1 \alpha &=& \partial_2 f/f \\
    \partial_2 \alpha &=& -\partial_1 f/f.
    \end{array}
\end{eqnarray}

A solution of Eq. (\ref{eq:psi}) only exists if $\nabla^2 \ln f = 0$, but this condition must, of course, be satisfied by bent helicoid minimal surfaces (if it was not, they could not have been constructed).
Eq. (\ref{eq:psi}) has a solution that is unique up to an overall constant. We can, for example, choose $\alpha$ at a single point on the surface, then integrate along a path to find $\alpha$ at any other point. Therefore, given any solution for $\alpha$ gives a solution to the Gauss-Codazzi-Mainardi system,
\begin{eqnarray}
    b_{1 1} &=& f \cos (\alpha + \theta), \nonumber \\
    b_{12} &=& f \sin (\alpha + \theta), \\
    b_{2 2} &=& -f \cos (\alpha + \theta) \nonumber,
\end{eqnarray}
where $\theta$ is a constant. Once a choice of $b_{i j}$ and $g_{i j}$ are made that are compatible with the Gauss-Codazzi-Mainardi system, the resulting surface is unique up to translations and rotations by Bonnet's theorem.
This shows that the Bonnet isometry of an associate family is the unique continuous transformation that only rotates the principal directions preserving both zero mean curvature and the isothermal coordinate system.

Notice that the matrix
\begin{eqnarray}
    b_i^{~j}  &=&  \frac{1}{2} \nabla^2 \ln \Omega \left( \begin{array}{cc}
       \cos (\alpha + \theta) & \sin (\alpha + \theta) \\
       \sin (\alpha + \theta) & -\cos (\alpha + \theta)
     \end{array}\right)
    % &=& \frac{1}{2} \nabla^2 \ln \Omega \left( \begin{array}{cc}
    %     \cos (\psi/2) & - \sin(\psi/2) \\
    %     \sin(\psi/2) & \cos(\psi/2)
    % \end{array}\right) \left(\begin{array}{cc}
    %     1 & 0 \\
    %     0 & -1
    % \end{array}\right) \left( \begin{array}{cc}
    %     \cos (\psi/2) &  \sin(\psi/2) \\
    %     -\sin(\psi/2) & \cos(\psi/2)
    % \end{array}\right). \nonumber
\end{eqnarray}
Thus, we can identify the principle curvatures as $\pm c$, where
\begin{equation}
    c = \pm \frac{1}{2} \left|\nabla^2 \ln \Omega \right|.
\end{equation}

This construction gives us the direction of the principal curvatures even when the surface is not simply connected, but it does not guarantee that the resulting surface remains closed as a function of $\theta$. Consider decomposing a minimal surface into simply-connected patches. Schwarz' theorem guarantees that each patch transforms uniquely as we change $\theta$ and, because of this, they can be made to agree whenever patches overlap by setting the Bonnet angles equal on each overlapping patch. However, there is no guarantee patches will overlap around any non-contractible loops. Hence, this leaves open the question of whether a Bonnet isometry exists globally, but it does show that no other isometry preserving $H=0$ could exist.

\subsection{Infinitesimal Isometries of minimal surfaces}

In this section, we present some calculations about the infinitesimal isometries of minimal surfaces. These isometries do not necessarily preserve mean curvature.

\subsubsection{Identities}
We collect a few useful identities for minimal surfaces. Let $e_+^i$ and $e_-^i$ be unit vectors associated with the largest and smallest principal curvatures of $\mathbf{X}_q(\rho,\phi)$. Since they are orthogonal,
\begin{eqnarray}
    e^+_i \epsilon^{i j} &=& e_-^j \nonumber \\
    e^-_i \epsilon^{i j} &=& e_+^j \nonumber
\end{eqnarray}
and so we can write
\begin{eqnarray}
    b^0_{i j} &=& c (e^+_i e^+_j - e^-_i e^-_j) \nonumber \\
   b^C_{i j} &=& c (e^+_i e^-_j + e^-_i e^+_j) \nonumber.
\end{eqnarray}

Notice that
\begin{equation}
    b_{i j} \epsilon^{j}_{~k} = \epsilon_{i j} b^{j}_{~k} = b^C_{i k}.
\end{equation}
and so
\begin{equation}
    b_{i j} {b_C}^{j}_{~k} = b_{i j} \epsilon^{j l} b_{lk} = c^2 \epsilon_{i k}.
\end{equation}
since the tensor is antisymmetric. Finally, this implies that
\begin{eqnarray}
    b_{i j} b_C^{i j} &=& 0 \nonumber \\
    b_{i j} g^{i j} &=& 0 \nonumber \\
    b_C^{i j} g_{i j} &=& 0 \nonumber
\end{eqnarray}

\subsubsection{Equation for infinitesimal isometries}
A generic infinitesimal isometry, $\delta \mathbf{X} = u^k \partial_k \mathbf{X} + \zeta \hat{\mathbf{N}}$, satisfies $D_i u_j + D_j u_i - 2 \zeta b_{i j} = 0$. This can be rewritten
\begin{eqnarray}
    D_i u^i &=& 0 \nonumber \\
    b_C^{i j} D_i u_j &=& 0 \\
    b^{i j} D_i u_j &=& b^{i j} b_{i j} \zeta = -2 K \zeta \nonumber
\end{eqnarray}
where $K$ is the Gaussian curvature.

The change in the curvature to first order is given by
\begin{eqnarray}
    \delta b_{i j} &=& \hat{\mathbf{N}} \cdot D_i D_j \delta \mathbf{X}
\end{eqnarray}
so, after some algebra, the change in the mean curvature is
\begin{eqnarray}
    2 \delta H &=& \nabla^2_g \zeta - 2 K \zeta,
\end{eqnarray}
where $\nabla^2_g$ is the Laplace-Beltrami operator on the surface.

The Hodge-Morrey-Friedrichs decomposition on an orientable manifold with boundaries has the form \cite{schwarz2006hodge}
\begin{equation}
    u_i = \partial_i \chi + \epsilon_{i}^{~k} \partial_k \eta + h_i,
\end{equation}
where $\chi$ is constant on each boundary component, $\eta$ is constant on each boundary component, and $b_i$, known as the harmonic field, satisfies $g^{i j} D_i h_i = 0$ and $\epsilon^{i j} D_i h_i = 0$.

On a minimal orientable ribbon, $\chi = c_1 \rho + c_2$ is the general solution. To show this, first,
\begin{equation}
    g^{i j} D_i u_j = \nabla^2_g \chi.
\end{equation}
The isometry equation implies $\nabla^2_g \chi = 0$.

The harmonic field can be further decomposed into a sum of the form
    \begin{equation}
        h_i = \tilde{h}_i + \epsilon_i^{~j} \partial_j \tilde{\eta}
    \end{equation}
where $\tilde{\eta}$ is a harmonic function, $\tilde{h}_i t^i = 0$ and $t^i$ is a vector tangent to the boundary.
Specializing to an isothermal coordinate system, it can be shown that
\begin{equation}
    (\partial_\rho^2 + \partial_\phi^2) \tilde{h}_i = 0.
\end{equation}
Since $\tilde{h}_\phi = 0$ on the boundary, the only solution is $\tilde{h}_\phi = 0$.
Therefore,
\begin{equation}
    u_i = c_1 \delta_{i \rho} + \epsilon_i^{~j} \partial_j \psi
\end{equation}
where $\delta_{i j}$ is the Kronecker delta and $\psi = \eta + \tilde{\eta}$.

This theorem does not strictly hold for non-orientable surfaces, however, we can apply it on the double cover of a bent helicoid by letting $\phi$ formally take values between $0$ and $4 \pi$. Letting $\chi_D$ and $\psi_D$ be the solutions on the double cover, we then reduce 
\begin{eqnarray}
    \chi(\rho,\phi) &=& \frac{1}{2} \left[ \chi_D(\rho,\phi) + \chi_D(-\rho,\phi+2 \pi) \right] = \textrm{constant}\\
    \psi(\rho,\phi) &=& \frac{1}{2} \left[ \psi_D(\rho,\phi) + \psi_D(-\rho,\phi+2 \pi) \right].
\end{eqnarray}

On a simply-connected patch, a divergence-free vector field can be written as
\begin{equation}
    u^i = \epsilon^{i j} \partial_j \psi.
\end{equation}
Then we obtain
\begin{equation}\label{eq:shear}
    b^{i j} D_i D_j \psi = 0.
\end{equation}
On a minimal surface, $D_i b^{i j} = 0$, so we can express this as the second-order hyperbolic equation,
\begin{equation}
    \partial_i \left( \sqrt{g} ~b^{i j} \partial_j \psi \right) = 0.
\end{equation}

An infinitesimal isometry that simultaneously preserves $H$ therefore satisfies the system,
\begin{eqnarray}
    0 &=& \nabla^2_g \zeta - 2 K \zeta,      \nonumber   \\
    -2 K \zeta &=& D_i \left( b_C^{i j} \partial_j \psi \right) \\
    0 &=& D_i \left( b^{i j} \partial_j \psi \right)   \nonumber
\end{eqnarray}

\subsubsection{The infinitesimal Bonnet isometry}

Assuming isothermal coordinates with conformal factor $\Omega$, the in-plane displacements produced by a change $\delta \theta$ along the Bonnet isometry are
\begin{equation}
\label{In-plane displacement}
u_i = \partial_\theta \mathbf{X} \cdot \partial_i\mathbf{X}_0 = -\sin \theta \left(\partial_i\mathbf{X}_0 \cdot \mathbf{X}_0\right) + \cos \theta \left(\partial_i\mathbf{X}_0 \cdot \mathbf{X}_C\right)
\end{equation}
Hence,
\begin{equation}
\label{vorticity}
    \nabla^2_g \psi = \epsilon^{ij}D_iu_j = \frac{1}{\Omega}\left(\partial_1u_2 - \partial_2u_1\right) = \frac{\cos \theta}{\Omega}\left(\partial_2\mathbf{X}_0\cdot \partial_1 \mathbf{X}_C - \partial_1\mathbf{X}_0\cdot \partial_2\mathbf{X}_C  \right)= - 2\cos \theta,
\end{equation}
where we have used the Cauchy-Riemann equations~(\ref{Cauchy-Riemann}). We set $\theta = 0$ since we are interested in the isometries of the reference bent helicoid, $\mathbf{X}_0 = \mathbf{X}_q$.

%Finally, we note that
%\begin{equation}
%    \triangle_g \psi( (-1)^q, \phi+2\pi) = \sigma \triangle_g \psi( \rho, \phi) = -2.
%\end{equation}
%Therefore, the Bonnet isometry is only compatible with a closed ribbon when $\sigma = 1$. This rationalizes the opening of the nonorientable ribbons when $q$ is odd, though it does not resolve the opening of $q=0$ and $q=2$ under a Bonnet isometry.

\section{Bent helicoid family}

\begin{figure}[h!]
    \centering
    \includegraphics[width=0.8\textwidth]{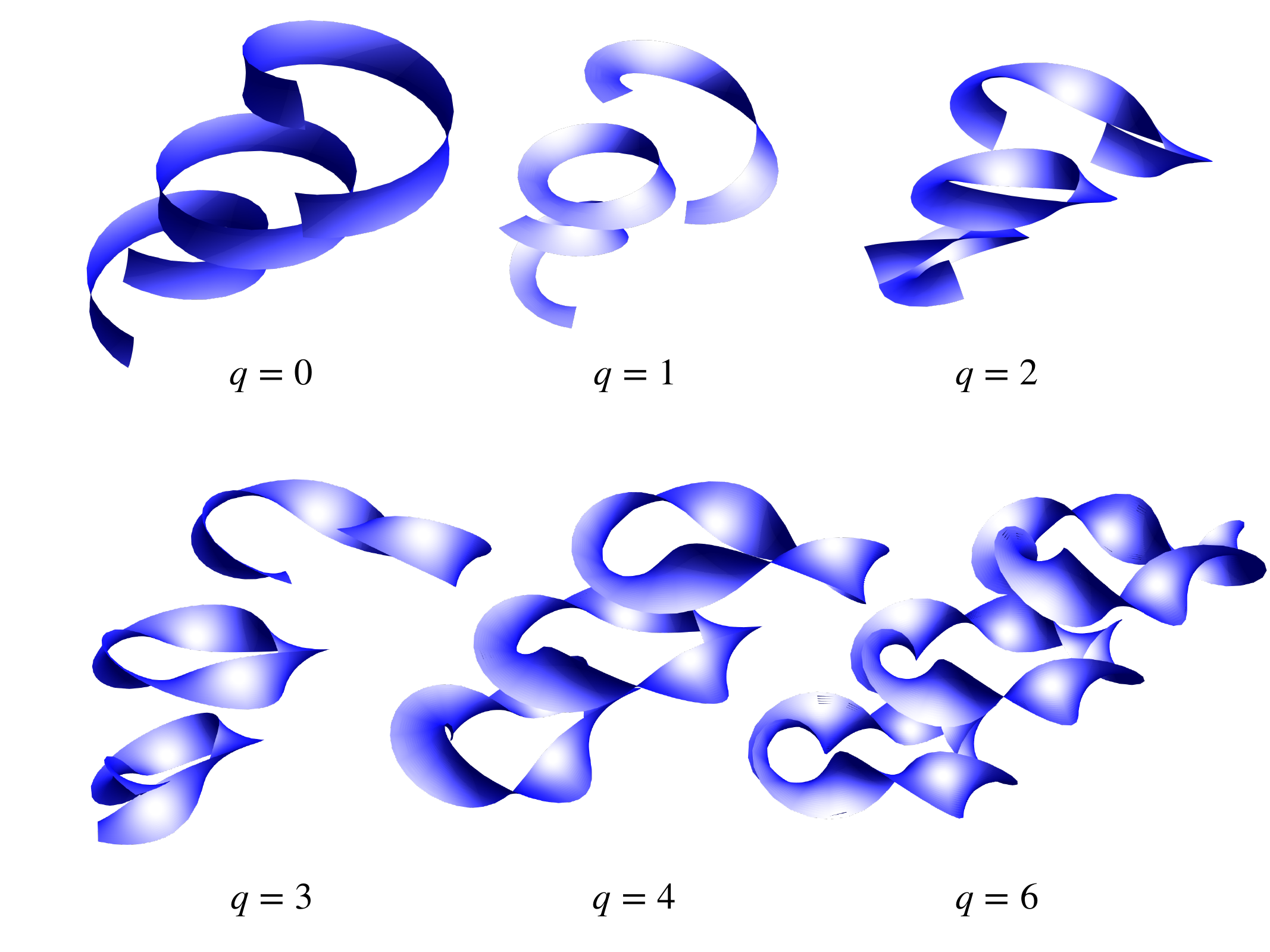}
    \caption{The bent helicoid family for several examples of $q$ half twists. Each member is a minimal surface and is isometric. Only helicoids with even $q > 2$ remain closed under the Bonnet isometry.}
    \label{fig:surfaces}
\end{figure}

\begin{figure}[h!]
    \centering
    \includegraphics[width=0.66\linewidth]{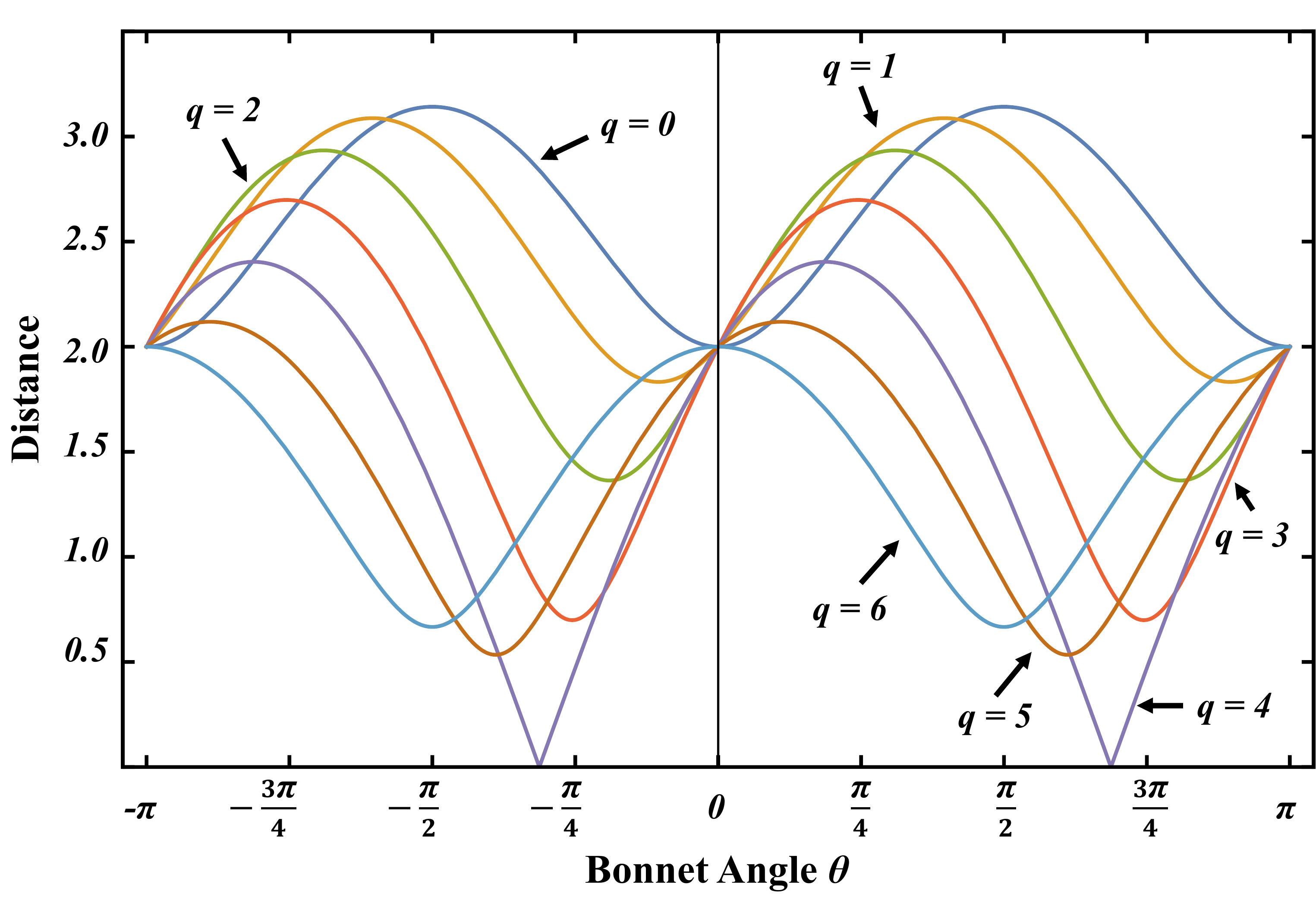}
    \caption{Distance between antipodal points $(0, \pm\pi/2)$ as a function of the Bonnet angle $\theta$ for the bent helicoids.}
    \label{fig: antipodal distance bent helicoids}
\end{figure}

\subsection{Parameterization of the bent helicoids}
The bent helicoid associate family of minimal surfaces \cite{meeks2007bending} with $q$ half twists can be constructed by solving a Bj\"orling problem (see Ref. \cite{lopez2018explicit}). In the Bj\"orling problem, one builds a minimal surface from an analytic base curve, which we take to be $\boldsymbol{\chi}(\phi) = \mathrm{R} (\sin \phi, \cos \phi, 0)^T$ where $\mathrm{R}$ is a constant and $\phi$ indicates a coordinate along the base curve. Consequently, $\chi'(\phi) = \mathrm{R} (\cos \phi, -\sin \phi, 0)^T$. We also prescribe a unit normal vector for the minimal surface along the base curve, $\hat{\mathbf{N}}_q(\phi) = \cos\left(q \phi/2\right) (\sin \phi, \cos \phi, 0)^T + \sin \left(q \phi/2\right) (0,0, 1)^T$. The entire associate family of Bj\"orling surfaces is given by analytically continuing $\boldsymbol{\chi}'$ and $\hat{\mathbf{N}}_q$ to the complex plane with $z = \phi + i \rho$,
\begin{equation}
    \label{eq. 1}
     \mathbf{X}_q = \text{Re} \left(e^{i \theta} \left[\boldsymbol{\chi}(z) - i \int \mathrm{d}z ~ \hat{\mathbf{N}}_q(z) \times \boldsymbol{\chi}'(z)  \right]\right)
\end{equation}
where $\theta$ is the Bonnet angle. When $\theta = 0$, Eq. (\ref{eq. 1}) describes a minimal surface for which $\boldsymbol{\chi}(\phi)$ and $\hat{\mathbf{N}}_q(\phi)$ is the shape and unit normal of the curve $\rho = 0$. Hence, along this curve, the tangent vectors to the closed bent helicoid, $\mathbf{X}_q(0, \phi; 0)$, are $\partial_\phi\mathbf{X}_q(0, \phi; 0) = \boldsymbol{\chi}'(\phi)$ and $\partial_\rho\mathbf{X}_q(0, \phi; 0) = \hat{\mathbf{N}_q}(\phi) \times \boldsymbol{\chi}'(\phi)$, and satisfy
\begin{eqnarray}
\partial_\rho\mathbf{X}_q(0, \phi + 2\pi) &=& (-1)^q \partial_\rho\mathbf{X}_q(0, \phi), \\
\partial_\phi\mathbf{X}_q(0, \phi + 2\pi) &=&  \partial_\phi\mathbf{X}_q(0, \phi). \nonumber
\end{eqnarray}

Our starting connected bent helicoids correspond to $\theta = 0$ in Eq.~(\ref{eq. 1}), and an explicit computation yields an expression for the cartesian coordinates for arbitrary $\theta$,
\begin{align}
       \label{eq:x}
    x_q(\rho, \phi; \theta)/R &= \frac{4 \cos \theta}{q^2 - 4}\left[- \cosh\left(\frac{q}{2} \rho\right) \sinh{\rho} \left(\sin (\phi) \sin \left(\frac{q}{2} \phi\right)+\frac{q}{2} \cos (\phi) \cos \left(\frac{q}{2} \phi\right)\right) + \right. \\
    & \left. 4 \cosh{\rho}\left(\left(q^2-4\right) \sin (\phi)+\sinh \left(\frac{q}{2} \rho \right) \left(\frac{q}{2} \sin (\phi) \sin \left(\frac{q}{2} \phi\right)+\cos (\phi) \cos \left(\frac{q}{2} \phi\right)\right)\right)\right] \nonumber \\
    & + 4 \frac{\sin \theta}{q^2 - 4}\left[\cosh{\rho} \cosh\left(\frac{q}{2} \rho\right)\left(\cos (\phi) \sin \left(\frac{q}{2} \phi\right)-\frac{q}{2} \sin (\phi) \cos \left(\frac{q}{2} \phi\right)\right) + \right. \nonumber \\
   & \left. \sinh{\rho} \left(\sinh \left(\frac{q}{2} \rho\right) (\sin (\phi) \cos \left(\frac{q}{2} \phi\right)-\frac{q}{2} \cos
   (\phi) \sin \left(\frac{q}{2} \phi\right))-4 \left(q^2-4\right) \cos
   (\phi)\right) \right] \nonumber, \\
       \label{eq:y}
   y_q(\rho, \phi; \theta)/R &= \frac{\cos \theta}{q^2 -1} \left[\cosh\left(q \rho\right) \sinh{\rho}\left(q \sin (\phi) \cos \left(q \phi\right)-\cos (\phi) \sin \left(q \phi\right)\right) +  \right. \\
   & \left. \cosh{\rho}\left(\left(q^2-1\right) \cos (\phi)+\sinh \left(\frac{q}{2} \rho\right) \left(\frac{q}{2} \cos (\phi) \sin \left(\frac{q}{2} \phi\right)-\sin (\phi) \cos \left(\frac{q}{2}
   \phi\right)\right)\right)\right] \nonumber \\
   &+ \frac{4\sin \theta}{q^2 - 4}\left[- \cosh{\rho}\cosh\left(\frac{q}{2} \rho\right)\left(\sin (\phi) \sin \left(\frac{q}{2} \phi\right)+\frac{q}{2} \cos (\phi) \cos \left(\frac{q}{2} \phi\right)\right) + \right. \nonumber \\
   & \left. \sinh{\rho}\left(\left(\frac{q^2}{4}-1\right) \sin (\phi)+\sinh \left(\frac{q}{2} \rho\right) \left(\frac{q}{2} \sin (\phi) \sin \left(\frac{q}{2} \phi\right)+\cos (\phi) \cos \left(\frac{q}{2} \phi\right)\right)\right) \right] \nonumber, \\
        \label{eq:z}
   z_q(\rho, \phi; \theta)/R &= - \frac{2\cos\theta}{q}\sinh\left(\frac{q}{2} \rho\right)\cos\left(\frac{q}{2} \phi\right)- \frac{2\sin \theta}{q}\cosh\left(\frac{q}{2} \rho\right) \sin\left(\frac{q}{2} \phi\right).
\end{align}

These parameterizations Eq. (\ref{eq:x} - \ref{eq:z}) are ill-defined when $q = 0$ and $q=2$. To construct the ribbons for these two cases, we return to the integral in Eq.~(\ref{eq. 1}). For $q=0$ we find
\begin{align}
     x_0(\rho, \phi; \theta)/R &=\cos\theta \cosh\rho\sin \phi - \sin\theta \sinh \rho \cos \phi\\
     y_0(\rho, \phi; \theta)/R &=\cos\theta \cosh\rho\cos\phi + \sin\theta\sinh\rho\sin\phi\\
     z_0(\rho, \phi; \theta)/R &=-\rho\cos\theta  - \phi\sin\theta ,
\end{align}
the formula for the catenoid-helicoid associate family.
 For $q = 2$,
\begin{align}
     x_2(\rho, \phi; \theta)/R &= \frac{\cos\theta}{4}\left(-\sinh (2 \rho ) \cos (2 \phi )+4 \cosh (\rho ) \sin (\phi )+2 \rho\right) \\
     & + \frac{\sin\theta}{4} \left(-4 \sinh (\rho ) \cos (\phi )-\cosh (2 \rho ) \sin (2 \phi )+2 \phi\right) \nonumber \\
     y_2(\rho, \phi; \theta)/R &=\frac{\cos\theta}{4} \left(\sinh2\rho\sin2\phi + 4 \cosh\rho\cos\phi\right) \\
     & -\frac{\sin\theta}{4}\left(-4 \sinh (\rho ) \sin (\phi )+\cosh (2 \rho ) \cos (2 \phi )+1\right) \nonumber \\
     z_2(\rho, \phi; \theta)/R &= -\sin \theta \cosh (\rho ) \sin (\phi )-\cos \theta  \sinh (\rho ) \cos (\phi )
\end{align}

The coordinates $(\rho,\phi)$ are isothermal, as we can see by computing the metric tensor
    \begin{equation}\label{eq:metric}
        g_{ij} dx^i dx^j \equiv \Omega(\rho,\phi) (d\rho^2 + d\phi^2),
    \end{equation}
where $\Omega = R^2 \left[\sinh (\rho) \sin \left(\frac{q}{2} \phi\right)+\cosh (\rho) \cosh \left(\frac{q}{2} \rho\right) \right]^2$. We assume that repeated indices are summed and the indices range over the coordinates $(\rho,\phi)$.

\subsection{Connectedness of the orientable bent helicoids under the Bonnet isometry}

To identify which ribbons become disconnected under the Bonnet isometry, it is sufficient to look at the ribbons' centerline $\rho = 0$ using the Bj\"orling parameterization in Eq.~(\ref{eq. 1}). The displacement between the endpoints of the centerline satisfies
\begin{align}
\Delta \mathbf{X}_q(\theta) &= \mathbf{X}_q(0, 2\pi; \theta) - \mathbf{X}_q(0, 0; \theta), \nonumber \\ 
&= \cos \theta \left[\boldsymbol{\chi}(2\pi) -\boldsymbol{\chi}(0) \right] + \sin \theta \int_0^{2\pi} \mathrm{ds} ~ \hat{\mathbf{N}}_q(s) \times \boldsymbol{\chi}'(s), \nonumber \\
&= \sin \theta \int_0^{2\pi} \mathrm{ds} ~ \hat{\mathbf{N}}_q(s) \times \boldsymbol{\chi}'(s).\label{Displacement}
\end{align}
The vanishing of the integral in Eq. (\ref{Displacement}) guarantees that the Bonnet isometry preserves the connectivity of the ribbon.

Despite the fact that Eq. (\ref{Displacement}) was derived only on the $\rho = 0$ line, it has broader validity. This is because the differential form,
\begin{equation}
    \omega(\rho,\phi) = du^i \epsilon_{i}^{~j} \partial_j \mathbf{X}(\rho, \phi),
\end{equation}
is closed for minimal surfaces, for which $ \nabla^2 \mathbf{X} = 0$. Therefore, $\int_\gamma \omega$ takes the same value on any non-contractible loop. Consequently,
\begin{equation}\label{form}
    \int_\gamma \omega = \int_0^{2 \pi} \mathrm{ds} ~ \hat{\mathbf{N}}(s) \times \gamma'(s).
\end{equation}

One can check by hand that the only orientable ribbons that fail this condition are $q = 0, 2$
\begin{eqnarray}
\label{resonances}
   \int_0^{2\pi} \mathrm{ds} ~ \hat{\mathbf{N}}_0(s) \times \boldsymbol{\chi}'(s) &=& \int_0^{2\pi} \mathrm{ds} ~ \left(0,0,-1\right)^T  = \left(0,0,-2\pi\right)^T,\\
   \int_0^{2\pi} \mathrm{ds} ~ \hat{\mathbf{N}}_2(s) \times \boldsymbol{\chi}'(s) &=& \int_0^{2\pi} \mathrm{ds} ~ \left(\sin^2s, \frac{\sin 2s}{2},-\cos s\right)^T = \left(\pi,0,0\right)^T, \nonumber
\end{eqnarray}
which we refer to as resonance conditions because of the appearance of Fourier-like modes in the integrands of Eqs.~(\ref{resonances}). Hence, the Bonnet isometry disconnects the $q = 0, 2$ bent helicoids but not higher even $q$.

It turns out that we can find minimal surfaces with $q=0$ and $q=2$ that do stay connected under the Bonnet isometry. Let $\boldsymbol{\chi}(\phi) = \left(\sin \phi, \cos \phi, 0\right)$, and the unit normals
\begin{equation}
\label{Unit normal bessel}
\mathbf{N}_n(\phi) = \left(-\sin \phi \sin \left[n \phi + a \sin \phi\right], -\cos \phi \sin \left[n \phi + a \sin \phi\right],   \cos \left[n \phi + a \sin \phi\right]\right),
\end{equation}
where $a$ is a real positive number and $n = \frac{q}{2}$ with even natural number $q$. This choice gives the tangent vector along the centerline
\begin{equation}
\label{tangent bessel}
\mathbf{N}_n(\phi) \times \boldsymbol{\chi}'(\phi) = \left(\sin \phi \cos \left[n \phi + a \sin \phi\right], \cos \phi \cos \left[n \phi + a \sin \phi\right], \sin \left[n \phi + a \sin \phi\right]\right).
\end{equation}
A direct computation shows that
\begin{eqnarray}
    \int _0^{2\pi} \mathrm{ds} ~ \mathbf{N}_{n}(s) \times \boldsymbol{\chi}'(s) = \left(-1\right)^{n + 1}\frac{2\pi n}{a}J_n(a)(0, 1, 0),
\end{eqnarray}
where $J_n$ is the n-th Bessel function of the first kind. We conclude that $q = n = 0$ produces a minimal surface with zero half-twists that stays connected under the Bonnet isometry for all values of $a$. The minimal surfaces for the other $n$ remain connected under the Bonnet isometry only if we choose $a$ to be any of the zeros of the n-th Bessel function. Thus, we can generate a minimal surface with two half-twists that stays connected under the isometry and many even-twisted minimal surfaces, with more than two twists, that do not.

We present the calculation of the integral for completeness. The integral for the $y$ Cartesian component reads
\begin{eqnarray}
    \int_0^{2\pi} \mathrm{ds} ~ \cos(ns + a \sin s)\cos s &=& \mathrm{Re} \int_0^{2\pi} \mathrm{ds} ~ \cos s ~ e^{ins} ~ e^{i a \sin s}, \nonumber \\
    &=& \mathrm{Re} \sum_{m = -\infty}^{m = \infty} J_m(a) \int_0^{2\pi} \mathrm{ds} ~ \cos s ~ e^{i s (n + m)},
\end{eqnarray}
where $\mathrm{Re}(\cdot)$ takes the real part of a complex number and we have used the Bessel-Fourier series expansion of $e^{i a \sin s}$. One can check that the only non-zero integrals in the sum require $n + m = \pm 1$. A computation using the recursion identities for the Bessel functions shows that
\begin{equation}
    \int_0^{2\pi} \mathrm{ds} ~ \cos(ns + a \sin s)\cos s = \pi \left[J_{1 - n}(a) + J_{-n -1}(a)\right] = \left(-1\right)^{n + 1}\frac{2\pi n}{a}J_n(a).
\end{equation}
A similar argument shows that the $x$ and $z$ Cartesian components of the integral vanish.\\

\subsection{Non-orientable minimal ribbons}
The integral in Eq.(\ref{Displacement}) is also non-zero for the non-orientable ribbons, but it is insightful to see this directly through the symmetry of the ribbons. We thank an anonymous referee for pointing out this argument.

Recall that, along the centerline $\rho =  0$, the tangent vectors of the closed non-orientable bent helicoids satisfy
\begin{eqnarray}
\label{tangent vectors ribbon}
\partial_\rho\mathbf{X}_q(0, \phi + 2\pi) &=& - \partial_\rho\mathbf{X}_q(0, \phi), \\
\partial_\phi\mathbf{X}_q(0, \phi + 2\pi) &=&  \partial_\phi\mathbf{X}_q(0, \phi). \nonumber
\end{eqnarray}
The Cauchy-Riemann equations (\ref{Cauchy-Riemann}) imply the additional symmetries for the tangent vectors of the conjugate surface, $\mathbf{X}_q^C$
\begin{eqnarray}
\label{tangent vectors conjugate ribbon}
\partial_\phi\mathbf{X}_q^C(0, \phi + 2\pi) &=& - \partial_\phi\mathbf{X}_q^C(0, \phi), \\
\partial_\rho\mathbf{X}_q^C(0, \phi + 2\pi) &=&  \partial_\rho\mathbf{X}_q^C(0, \phi). \nonumber
\end{eqnarray}
A direct computation using these symmetries and the Cauchy-Riemann equations shows that
\begin{equation}
\label{relation between symmetries}
\partial_\phi \left[\mathbf{X}_q^C(0, \phi + 2\pi) - \mathbf{X}_q^C(0, \phi)\right] = -2\partial_\phi\mathbf{X}_q^C(0, \phi) = -2\partial_\rho\mathbf{X}_q(0, \phi).
\end{equation}
The conjugate surface, $\mathbf{X}_q^C$ is explicitly given by \cite{dierkes2010minimal}
\begin{equation}
\label{Conjugate surface}
\mathbf{X}_q^C(\rho, \phi) = \int_0^\phi \partial_\rho \mathbf{X}_q(\rho, \tilde{\phi}) ~ \mathrm{d\tilde{\phi}} ~ -  \int_0^\rho \partial_\phi \mathbf{X}_q(\tilde{\rho}, 0) ~ \mathrm{d\tilde{\rho}}.
\end{equation}
Integrating Eq.(\ref{relation between symmetries}) from $0$ to $\phi$, and evaluating Eq.~(\ref{Conjugate surface}) along the centerline, $\rho = 0$, shows that
\begin{equation}
\label{symmetry conjugate surface non-orientable}
\mathbf{X}_q^C(0, \phi + 2\pi) = - \mathbf{X}_q^C(0, \phi) + \int_0^{2\pi} \partial_\rho \mathbf{X}_q(0, \tilde{\phi}) ~ \mathrm{d\tilde{\phi}},
\end{equation}
where the integral in the last expression is the integral in Eq. (\ref{Displacement}). Therefore, for odd $q$, the conjugate bent helicoids are always disconnected.

Finally, we note that this argument applies generally to non-orientable minimal surfaces. If $\mathbf{X}^0$ is non-orientable, there exists a non-contractible loop, $\alpha$, over which the unit normal of the surface, $\hat{\mathbf{N}}^0$, flips sign after parallel transporting around the loop, $\hat{\mathbf{N}}^0 \mapsto - \hat{\mathbf{N}}^0$. Using the Bj\"orling construction, a small neighborhood around $\alpha$ can be reconstructed in isothermal coordinates $(\rho, \phi)$ using the loop $\alpha(\phi)$ and the unit normal $\hat{\mathbf{N}}^0(\phi)$. The resulting surface, $\mathbf{X}_\text{b}^0$ (b for Bj\"orling) is a ribbon whose tangent vectors along the centerline, $\partial_\phi\mathbf{X}_\text{b}^0 (0, \phi) = \alpha'(\phi)$ and $\partial_\rho\mathbf{X}_\text{b}^0(0, \phi) = \hat{\mathbf{N}}^0(\phi) \times \alpha'(\phi)$, satisfy the same symmetries as the tangent vectors of the non-orientable bent helicoids, Eqs.~(\ref{tangent vectors ribbon}), by construction. The displacement of the endpoints on the centerline is still given by Eq.~(\ref{Displacement}) because the centerline is a closed loop. The Cauchy-Riemann equations Eqs.~(\ref{Cauchy-Riemann}) hold as well in this region because the ribbon is parameterized in isothermal coordinates. Thus, the same argument as the one used for the non-orientable bent helicoids shows that the Bonnet isometry of the ribbon contained in $\mathbf{X}^0$, $\mathbf{X}_\text{b} = \cos \theta \mathbf{X}^0_\text{b} + \sin \theta \mathbf{X}^{0,C}_\text{b}$, disconnects the ribbon. Hence, the original surface, $\mathbf{X}^0$, also becomes disconnected.

\subsection{Symmetry of isometries}
Suppose that $\psi[ (-1)^q \rho, \phi+2\pi] = \sigma \psi(\rho,\phi)$ for some sign $\sigma = \pm 1$. Then
\begin{equation}
    \begin{array}{lcl}
    u_\rho( \rho,\phi) &=& \partial_\phi \psi(\rho,\phi) \\
    u_\phi( \rho,\phi) &=& -\partial_\rho \psi(\rho,\phi)
    \end{array}.
\end{equation}
Therefore, we obtain
\begin{equation}
    \begin{array}{lcl}
    u_\rho[ (-1)^q \rho,\phi+2\pi] &=& \sigma u_\rho( \rho,\phi) \\
    u_\phi[ (-1)^q \rho,\phi+2\pi] &=& -\sigma u_\phi( \rho,\phi)
    \end{array}.
\end{equation}
Since $\zeta = b^{i j} D_i u_j/(4 K)$, we can use $u_i$ to obtain the symmetry of $\zeta$, $\zeta[ (-1)^q \rho,\phi+2\pi] = \sigma \zeta(\rho,\phi)$.

Finally, we consider the symmetry of $\delta \mathbf{X}(\rho,\phi) = u^i \partial_i \mathbf{X} + \zeta \hat{\mathbf{N}}$,
\begin{equation}
    \delta \mathbf{X}[ (-1)^q \rho,\phi + 2 \pi] = (-1)^q \sigma \delta \mathbf{X}(\rho,\phi)
\end{equation}
Therefore, any isometry that preserves the connectivity of a ribbon must have $\sigma = (-1)^q$.

\subsection{Infinitesimal isometries of the catenoid}
To find the infinitesimal isometries of the catenoid, we express Eq. (\ref{eq:shear}) in isothermal coordinates,
\begin{equation}
    \cosh^2 \rho \partial_\rho \left( \frac{\partial_\rho \psi}{\cosh^2 \rho} \right) - \partial_\phi^2 \psi = 0.
\end{equation}
Using the expansion $\psi = \sum_m \psi_m(\rho) e^{i m \phi}$, we have to solve the system of ODEs,
\begin{equation}
    \cosh^2 \rho \partial_\rho \left( \frac{\partial_\rho \psi_m}{\cosh^2 \rho} \right) + m^2 \psi_m = 0
\end{equation}
which has a solution
\begin{equation}
\label{Infinitesimal isometries catenoid}
    \psi_m = a_m \cosh \rho P_1^{~\sqrt{1-m^2}}(\tanh \rho) + b_m \cosh \rho Q_1^{~\sqrt{1-m^2}}(\tanh \rho),
\end{equation}
where $P_l^{~n}(x)$ and $Q_{l}^{~n}(x)$ are the associated Legendre polynomials. Note that they can be analytically continued to complex $n$.

We can also compute the circulation of this solution, $-\triangle \psi = \chi$, to see that
\begin{equation}
    \chi = -\frac{1}{\cosh^2 \rho} \sum_{m} e^{i m \phi} \left( m^2 \psi_m(\rho) - \psi_m''(\rho) \right).
\end{equation}
Note that $\int d\phi ~ \chi(\rho,\phi) = 4 b_0 \tanh \rho$ showing that the circulation along the centerline, $\rho=0$ has at least two zeros. The Bonnet isometry, therefore, cannot have zero shear strain while maintaining the periodicity of the catenoid.

\begin{figure}
   \centering
   \includegraphics[width=1 \textwidth]{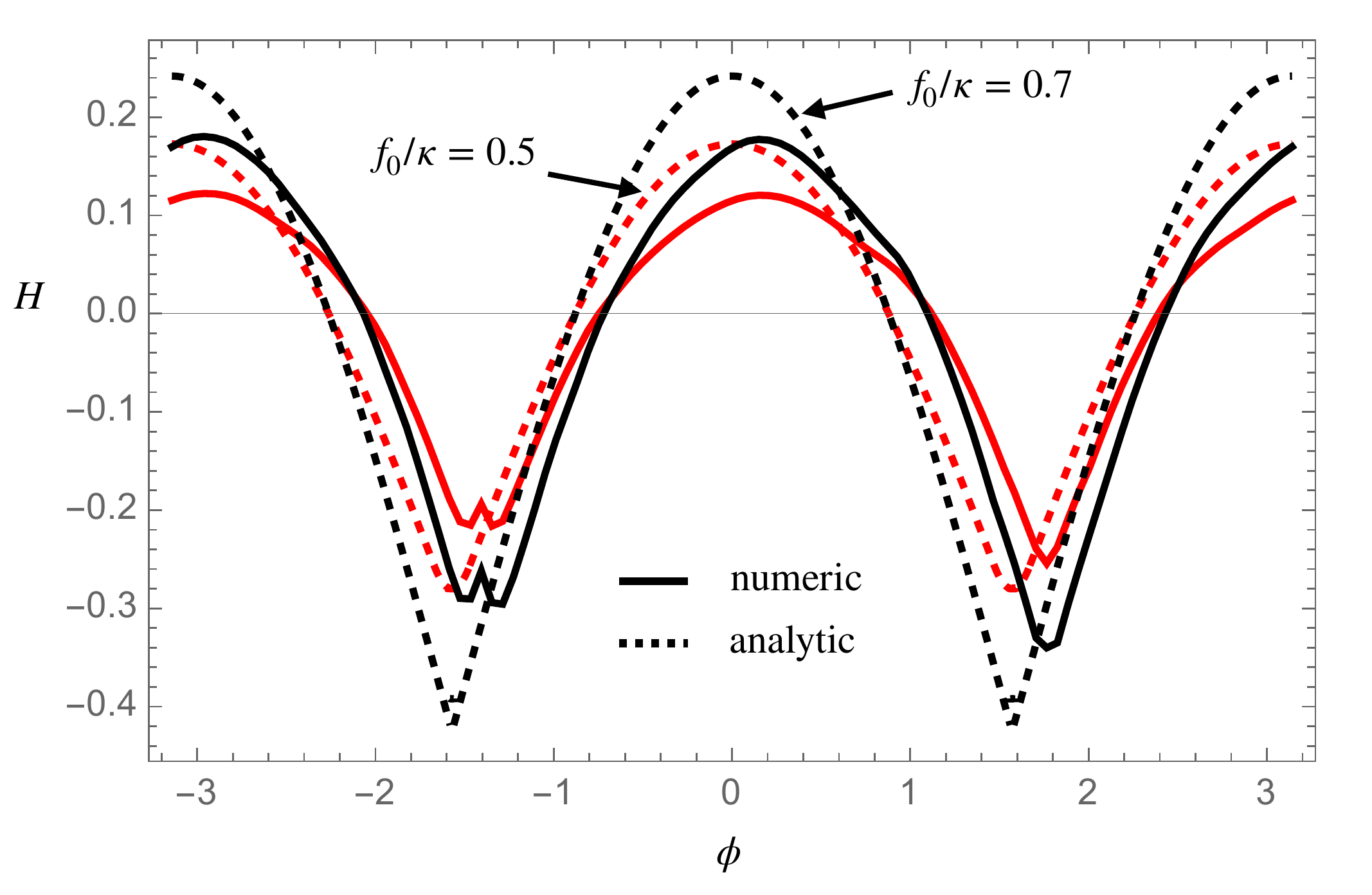}
   \caption{The mean curvature of the catenoid ($q=0$) along the center line ($\rho=0$) for $f_0/\kappa = 0.5$ and $0.7$ The analytic solution is shown as a dashed curve and the simulations are shown as solid curves.}
   \label{fig: catenoid solution}
\end{figure}

Although Eq.~(\ref{Infinitesimal isometries catenoid}) provides an analytical expression for all infinitesimal isometries of the catenoid, it is easier and more insightful to consider the series expansion, $\psi = \sum_n \psi_n(\phi)\rho^n$. The bending energy is
\begin{equation}
    E = \int d\phi ~ \left\{\kappa w \left[ \left(\psi_1 + \psi_1''\right)' \right]^2 + \frac{1}{3} \kappa w^3 \left[ (\psi_0 + \psi_0'')''' \right]^2 + \frac{1}{3} \kappa w^3 (\psi_1 + \psi_1'')' (\psi_1 + 2 \psi_1' + \psi_1'')' + \mathcal{O}(w^5)\right\}, \label{eq:q0energy}
\end{equation}
and assuming a force directed along the normal of magnitude $\delta(\rho) f(\phi)$, the work done is
\begin{equation}
    W = -\int d\phi ~ f(\phi) \psi_1'(\phi).
\end{equation}
We balance these two energies, as our numerical simulations of the catenoid suggest that the stretching energy is negligible. The leading order term, $\psi_0$, satisfies
\begin{equation}
\label{psi_0}
\left(\psi_0 + \psi_0''\right)''' = 0,
\end{equation}
from which we obtain the general solution $\psi_0(\phi) = c_0 + c_1 \phi + c_2 \phi^2 + c_3 \cos \phi + c_4 \sin \phi$. The constant $c_0$ does not correspond to a deformation; hence, we set it to zero. All other constants except $c_2$ correspond to Euclidean motions, but $c_2$ describes a deformation for which $\epsilon^{i j} D_i u_j = c_2$. We therefore interpret it as an infinitesimal Bonnet isometry. However, since $\psi_0'(\phi)$ is not periodic unless $c_2 = 0$, we see that the Bonnet isometry is, indeed, suppressed. The equilibrium equation for $\psi_1(\phi)$ is
\begin{equation}\label{eq:q=0 equation}
    \mathrm{A} + \frac{3f(\phi)}{2\kappa w(3 + w^2) } - \psi_1' -2 \psi_1^{(3)} - \psi_1^{(5)} = 0
\end{equation}
to lowest order in $w$ and $\mathrm{A}$ is a constant. To simulate pulling, we let $f(\phi) = f_0/\Delta w$ in two bands of width $\Delta w$ centered on $\pm \pi/2$. Assuming $\psi_1$ is continuous and periodic, the resulting solution is shown in Fig.~(\ref{fig: catenoid solution}). We note that the solution shows good agreement with
numerics with no fitting parameters. Unfortunately, for $q \ne 0$ the functions $\psi_0(\phi)$ and $\psi_1(\phi)$ become mixed at every order, and we are unable to find a simple solution.

\begin{comment}
\subsection{Constant rotation of in-plane deformations of the Bonnet isometry}

Assuming isothermal coordinates with conformal factor $\Omega$, the in-plane displacements produced by a change $\delta \theta$ along the Bonnet isometry are
\begin{equation}
\label{In-plane displacement}
u_i = \partial_\theta \mathbf{X} \cdot \partial_i\mathbf{X}_0 = -\sin \theta \left(\partial_i\mathbf{X}_0 \cdot \mathbf{X}_0\right) + \cos \theta \left(\partial_i\mathbf{X}_0 \cdot \mathbf{X}_C\right)
\end{equation}
Hence,
\begin{equation}
\label{vorticity}
    \Delta_{g} \psi = \epsilon^{ij}D_iu_j = \frac{1}{\Omega}\left(\partial_1u_2 - \partial_2u_1\right) = \frac{\cos \theta}{\Omega}\left(\partial_2\mathbf{X}_0\cdot \partial_1 \mathbf{X}_C - \partial_1\mathbf{X}_0\cdot \partial_2\mathbf{X}_C  \right)= - 2\cos \theta,
\end{equation}
where $\Delta_{g}$ is the Laplace-Beltrami operator of the minimal surface, and we have used the Cauchy-Riemann equations~(\ref{Cauchy-Riemann}). We set $\theta = 0$ since we are interested in the isometries of the reference bent helicoid, $\mathbf{X}_0 = \mathbf{X}_q$.

Finally, we note that
\begin{equation}
    \triangle_g \psi( (-1)^q, \phi+2\pi) = \sigma \triangle_g \psi( \rho, \phi) = -2.
\end{equation}
Therefore, the Bonnet isometry is only compatible with a closed ribbon when $\sigma = 1$. This rationalizes the opening of the nonorientable ribbons when $q$ is odd, though it does not resolve the opening of $q=0$ and $q=2$ under a Bonnet isometry.
\end{comment}

\section{Numerics}

\subsection{Discrete twisted ribbons}
\begin{figure}[h!]
    \centering
    \includegraphics[width=0.95\textwidth]{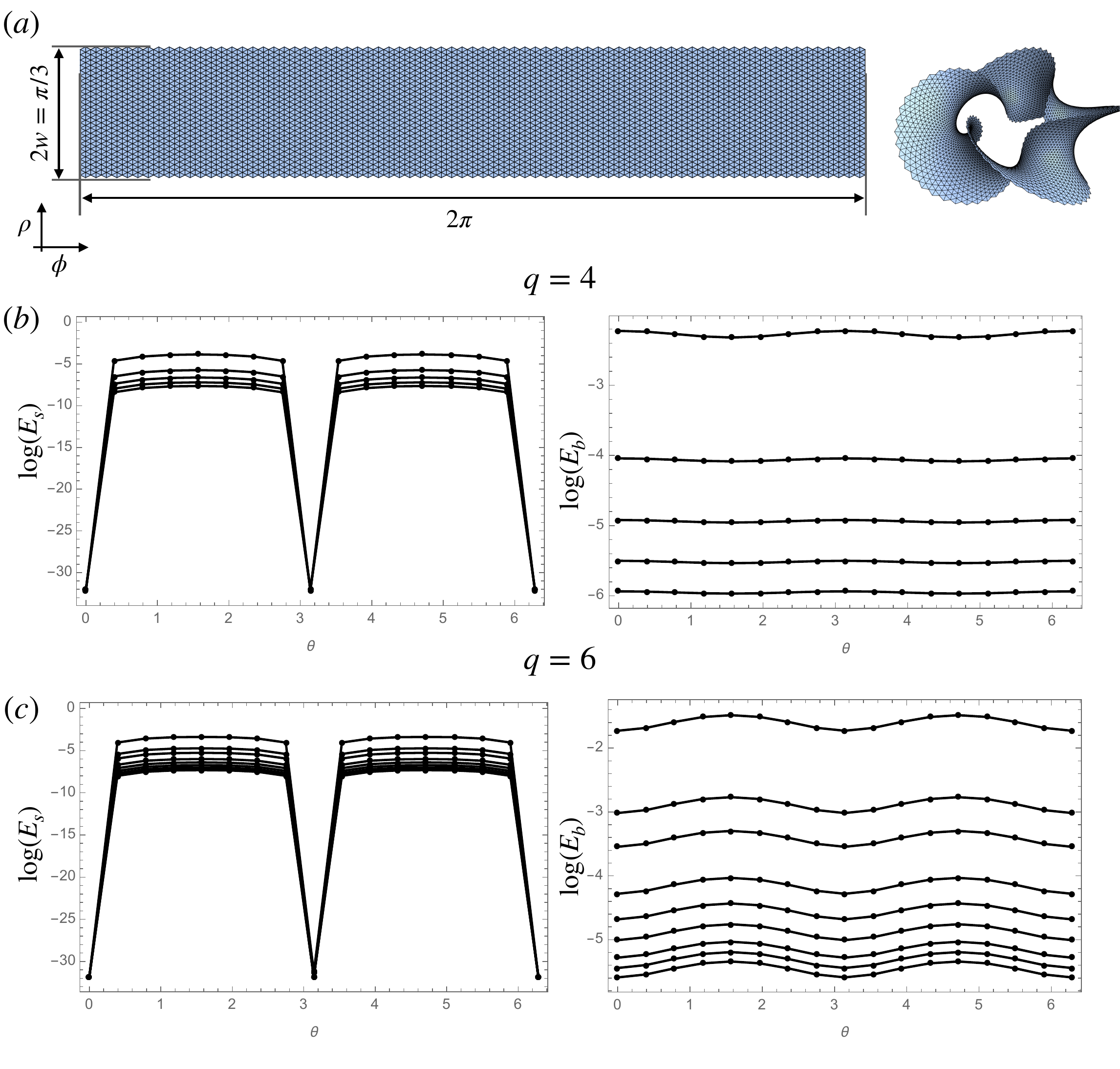}
    \caption{(a) An example triangular grid and bent helicoid ($q=4$ with side length $\pi/75$). The elastic and bending energies, for (b) $q=4$ and (c) $q=6$, as a function of Bonnet angle for side lengths $\pi/15$, $\pi,45$, $\pi/75$, $\pi/105$, and $\pi/135$ from top to bottom. The vertical scale is logarithmic.}
    \label{fig:grid}
\end{figure}

The ribbon is modeled as a triangular mesh (Fig. \ref{fig:grid}) whose vertices have coordinates in $(\rho,\phi)$. The equilibrium length of the edge joining vertex $i$ and vertex $j$, $\bar{l}_{i j}$, is determined by mapping the rectangular grid to $\mathbf{X}_q$.

For the stretching energy, we use
\begin{equation}
    E_s = \frac{1}{2} \sum_{\langle i, j \rangle} a_{ij} \left( l_{i j}/\bar{l}_{i j} - 1 \right),
\end{equation}
summed over all pairs of edges $\langle i, j \rangle$. Here, $a_{i j}$ denotes the $1/3$ the area of the triangles associated with the edge between vertex $i$ and $j$. To test the convergence of this energy, we compute the stretching energy of a flat rectangular sheet biaxially expanded by 10\%. Then $E_s/A = 0.01$ where $A$ is the area of the sheet before stretching. We further tested that the elastic energy increases quadratically with strain.

For the bending energy, we use
\begin{equation}
    E_b = \frac{1}{2} \kappa \sum_{i} a_i H_i^2,
\end{equation}
where $H_i$ is the vertex mean curvature, $a_i$ is $1/3$ the areas of the faces sharing the edge, and the sum is over all interior vertices. To compute $H_i^2$, we first orient the faces surrounding vertex $i$, then compute the change in area when displacing the vertex using the graph Laplacian formula, and finally square the result.

For $q=4$ and $q=6$, we plot the elastic stretching and bending energies as a function of Bonnet angle in Fig. \ref{fig:grid}. While the Bonnet isometry is not strictly a symmetry of the ribbons, it appears to converge (albeit slowly) to zero as the resolution increases.

\subsection{Cyclic loading}

\begin{figure}[h!]
    \centering
    \includegraphics[width=0.7\textwidth]{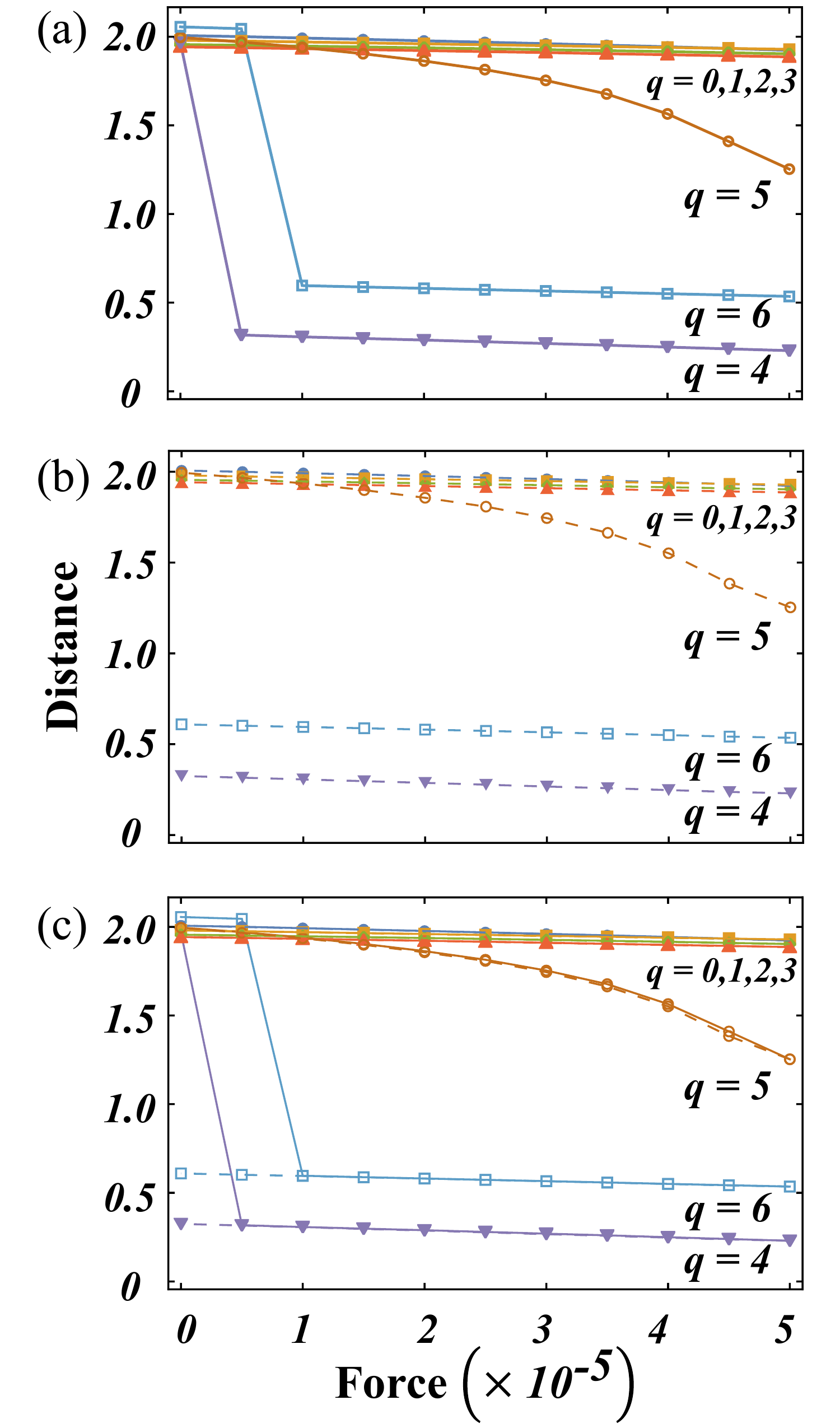}
    \caption{Distance between antipodal points $(0,\pm \pi/2)$ as a function of the applied force, $f_0$, under cyclic loading. (a) Increasing force. (b) Decreasing force. (c) The full trajectory under cyclic forcing shows that the $q= 4, 6$ bent helicoids are hysteretic while the rest are elastic.}
    \label{fig:loading}
\end{figure}

We performed a cyclic loading experiment where we ramped up the dimensionless force, $f_0$, and decreased it back to zero for the bent helicoids with $q$ up to 6 (Fig.~\ref{fig:loading}). The topologically rigid ribbons ($q = 0, 1, 2, 3, 5$) describe a nearly perfect reversible deformation, confirming that their response is elastic. However, the ultrasoft ribbons ($q = 4, 6$) do not return to their starting equilibrium length.

\subsection{Distance plateau and isometric limit}

\begin{figure}[h!]
    \centering
    \includegraphics[width=1\linewidth]{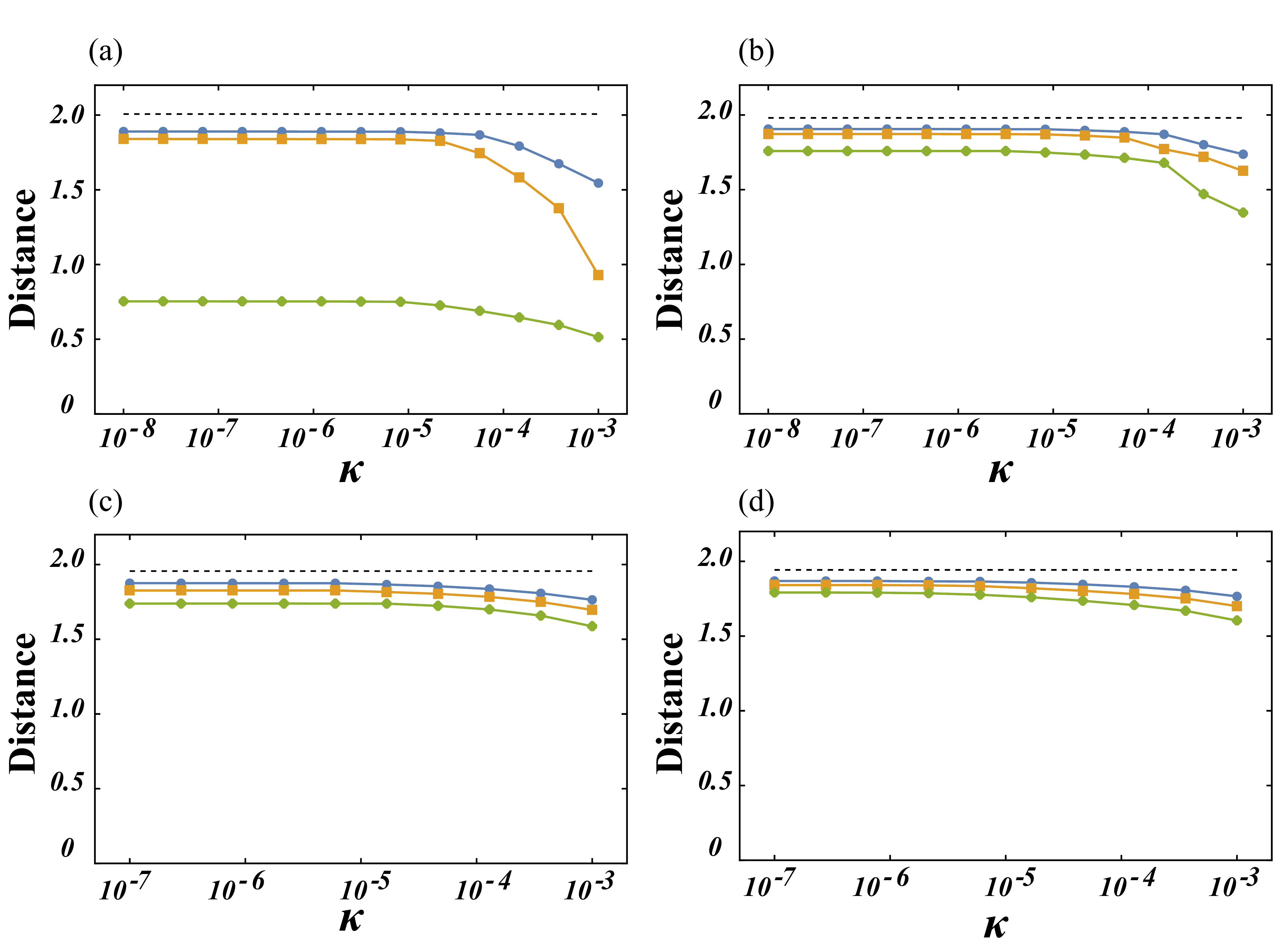}
    \caption{Distance between antipodal points $(0,\pm \pi/2)$ as a function of the bending rigidity $\kappa$ for fixed force ratio $f_0/\kappa$ for (a) $q = 0$, (b) $q = 1$, (c) $q = 2$, and (d) $q = 3$ bent helicoids. The dotted line is the initial distance before pinching. \emph{Diamond markers}: $f_0/\kappa = 1.0$, \emph{square markers}: $f_0/\kappa = 0.7$, \emph{circle markers}: $f_0/\kappa = 0.5$. All curves plateau towards the initial distance as $\kappa$ decreases.}
    \label{fig: Dist plateau}
\end{figure}

We plot the distance between antipodal points as a function of the bending rigidity at fixed force ratio (Fig.~\ref{fig: Dist plateau}). The plateau in the curves indicates that the stretching energy becomes subdominant, reaching the isometric limit as $f_0/\kappa$ decreases, and the final distance is
determined by a balance between the force and bending energy.

\subsection{Deformations of the ultrasoft bent helicoids}

\begin{figure}[h!]
    \centering
    \includegraphics[width=0.9\linewidth]{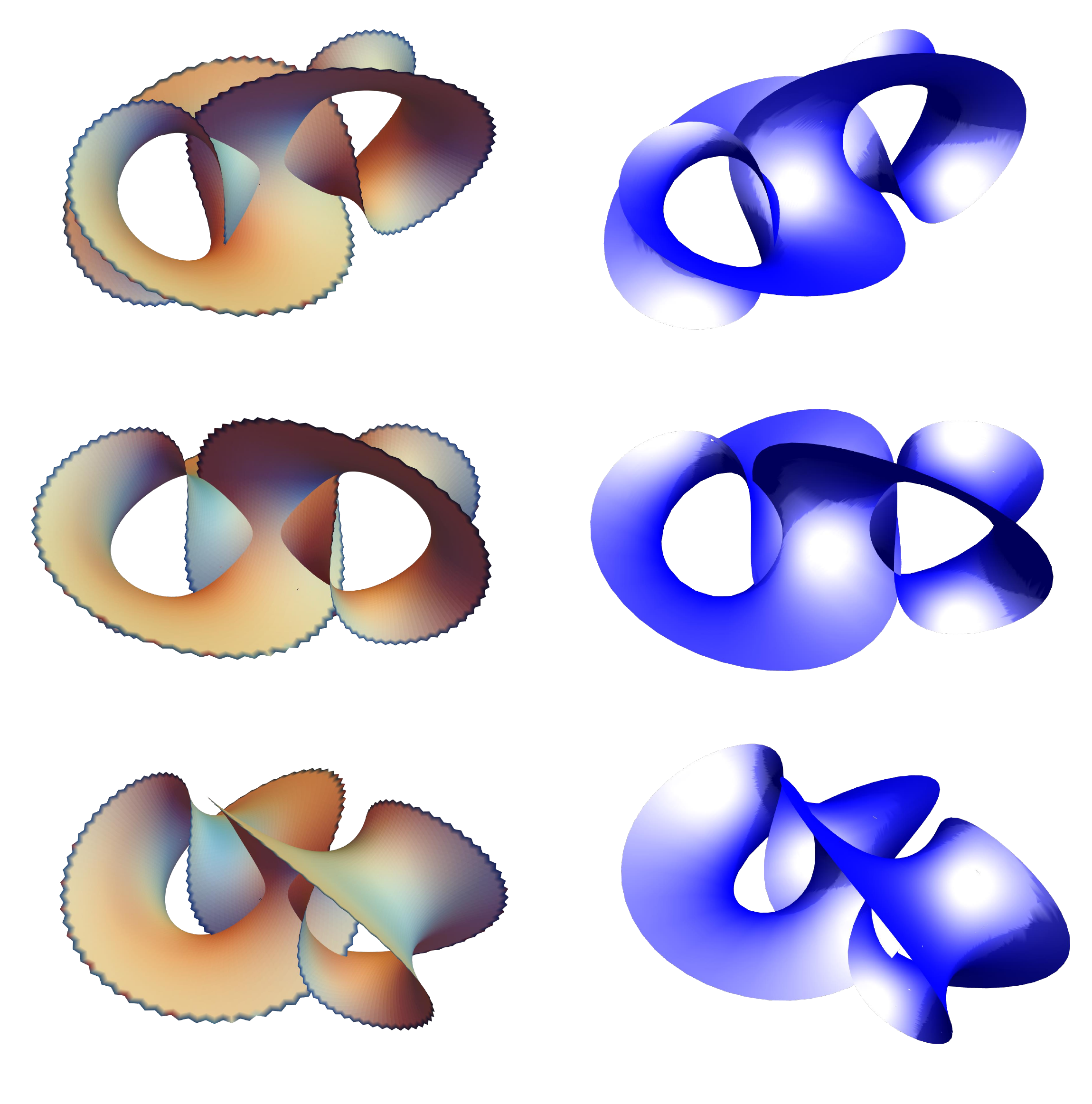}
    \caption{Deformation of the $q = 4$ bent helicoid. \emph{Left column}: Numerical ribbon colored by mean curvature. \emph{Right column}: Bonnet isometry with $\theta = -0.89$.}
    \label{fig:q4ultrasoft}
\end{figure}

\begin{figure}[h!]
    \centering
    \includegraphics[width=0.9\linewidth]{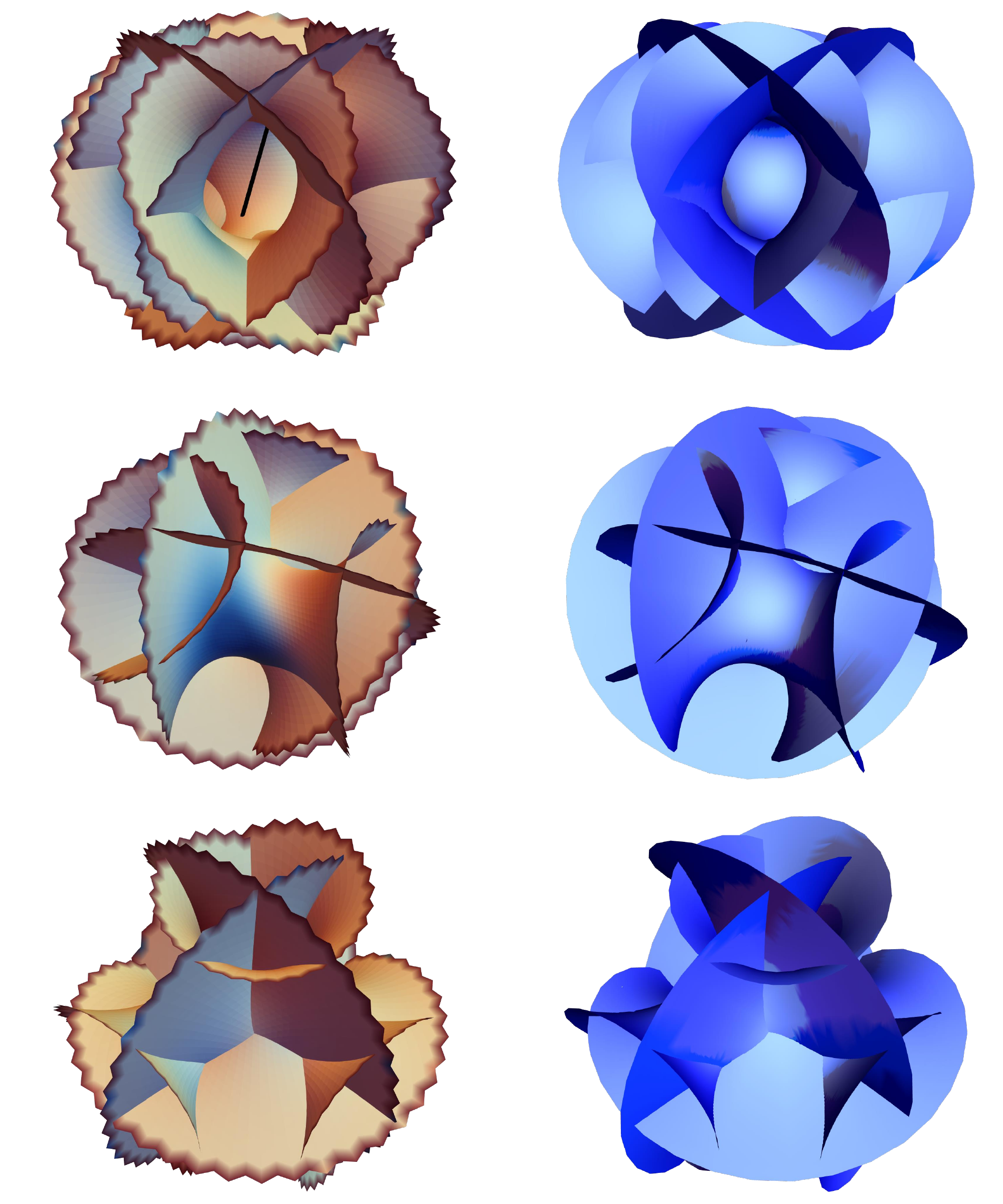}
    \caption{Deformation of the $q = 6$ bent helicoid. \emph{Left column}: Numerical ribbon colored by mean curvature. \emph{Right column}: Bonnet isometry with $\theta = 1.50$. }
    \label{fig:q6ultrasoft}
\end{figure}

\begin{figure}[h!]
    \centering
    \includegraphics[width=1\linewidth]{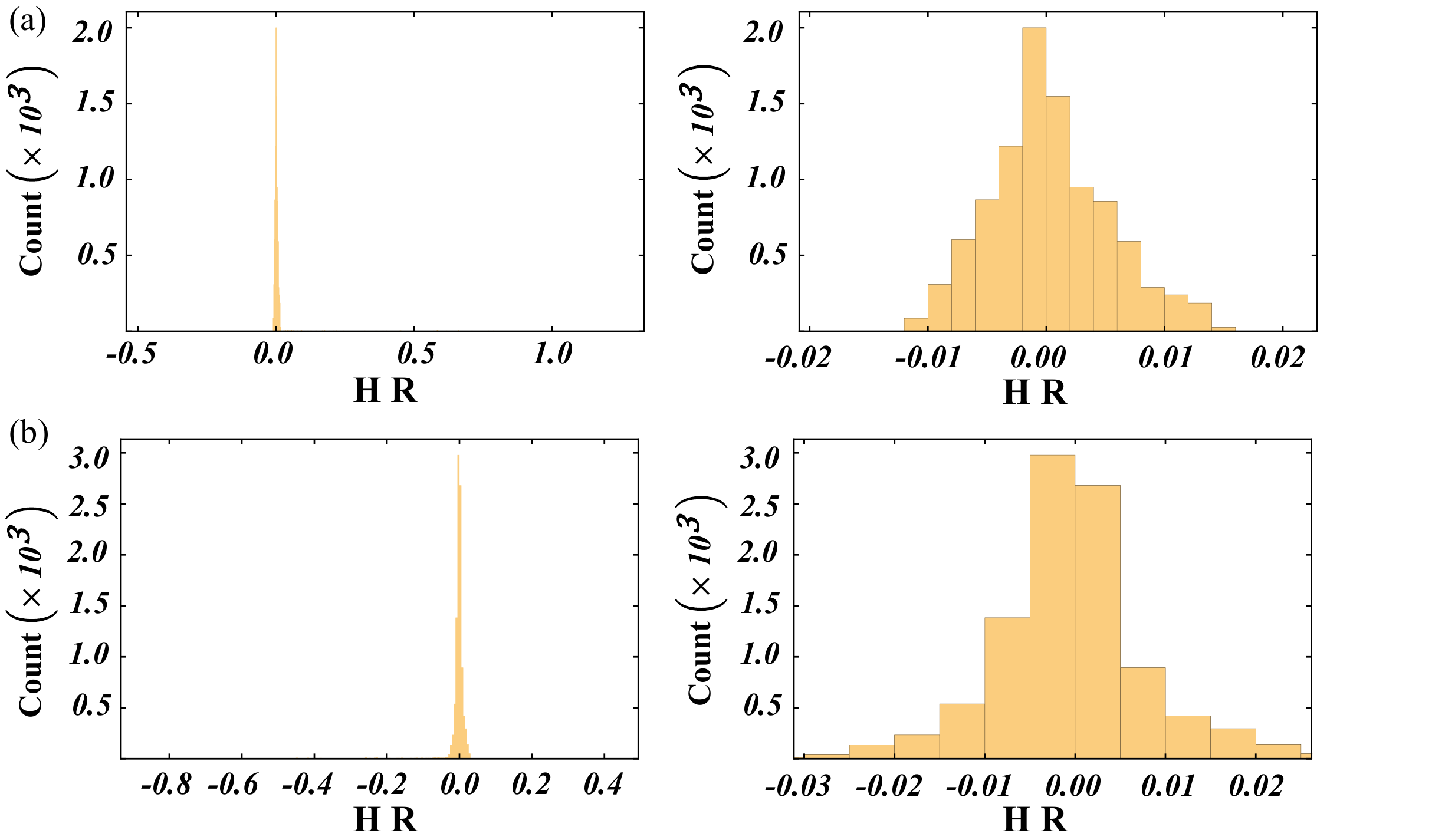}
    \caption{Dimensionless discrete mean curvature distributions, $\mathrm{H} \mathrm{R}$ where $\mathrm{H}$ is the mean curvature of the deformed ribbon and $\mathrm{R}$ is the radius of the centerline used to solve the Bj\"orling problem, of (a) $q = 4$ and (b) $q = 6$ bent helicoids. \emph{Left}: Full distribution. \emph{Right}: Zoom-in of the distribution around $0$ for which the mean and standard deviation are calculated. Numerical ribbons' color maps are based on these values. }
    \label{fig:Mean curvature histograms}
\end{figure}

Although we know the $q = 4, 6$ bent helicoids deform along a soft mode because there is almost no change in energy due to the pinching, we can also check that they are in reasonable agreement with a member of their corresponding associate families (Fig.~\ref{fig:q4ultrasoft} and Fig.~\ref{fig:q6ultrasoft}). The antipodal distances between the pinched surfaces and the analytical surfaces are also consistent. Fig.~\ref{fig:Mean curvature histograms} shows the vertex discrete mean curvature of the deformed ribbons. The minimum and maximum values are $-0.505777$ and $1.2932$ for $q = 4$, and $-0.901777$ and $0.4644$ for $q = 6$. However, the histograms show that these are outliers, most likely associated with the edges of the numerical ribbons. Most of the distribution is concentrated around $0$ for both cases, with values of  $0.014 \pm 0.024$ for $q = 4$ and $-0.0098 \pm 0.0299$ for $q = 6$, indicating both ribbons stay close to the zero mean curvature condition after the pinching. The ribbons are colored using values in the zoom-in distributions. These results indeed suggest that the ultrasoft ribbons deformed along the Bonnet isometry.

\bibliography{Supp_references}